\documentclass[lettersize,journal]{IEEEtran}
\usepackage{amsmath,amsfonts}
\usepackage{algorithmic}
\usepackage{algorithm}
\usepackage{array}
\usepackage{textcomp}
\usepackage{stfloats}
\usepackage{url}
\usepackage{verbatim}
\usepackage{graphicx}
\usepackage{siunitx}
\usepackage{comment}
\usepackage{multirow}
\usepackage{xcolor} 
\usepackage{booktabs}        
\usepackage{array}         
\usepackage{makecell}
\usepackage{hyperref}
\usepackage{subfigure}

\def\BibTeX{{\rm B\kern-.05em{\sc i\kern-.025em b}\kern-.08em
    T\kern-.1667em\lower.7ex\hbox{E}\kern-.125emX}}
\usepackage{balance}
\usepackage{acro}

\DeclareAcronym{3G}{short = 3G , long = third generation}
\DeclareAcronym{3GPP}{short = 3GPP , long = Third Generation Partnership Project}
\DeclareAcronym{4G}{short = 4G , long = fourth generation}
\DeclareAcronym{5G}{short = 5G , long = fifth generation}
\DeclareAcronym{6G}{short = 6G , long = sixth generation}

\DeclareAcronym{AoA}{short = AoA , long = angle of arrival}
\DeclareAcronym{AoD}{short = AoD , long = angle of departure}
\DeclareAcronym{AS}{short = AS , long = acceleration structure}
\DeclareAcronym{AGR}{short = AGR , long = adaptive grid refinement}

\DeclareAcronym{BBU}{short = BBU ,  long = baseband unit}
\DeclareAcronym{BS}{short = BS ,  long = base station}

\DeclareAcronym{CN}{short = CN , long = core network}
\DeclareAcronym{CPU}{short = CPU , long = central processing unit}
\DeclareAcronym{CH}{short = CH , long = closest hit}
\DeclareAcronym{CSI}{short = CSI , long = channel state information}
\DeclareAcronym{CDF}{short = CDF , long = cumulative distribution function}
\DeclareAcronym{CI}{short = CI , long = close-in}

\DeclareAcronym{EIRP}{short = EIRP , long = effective isotropic radiated power}
\DeclareAcronym{eMBB}{short = eMBB , long = enhanced mobile broadband}
\DeclareAcronym{EMF}{short = EMF , long = electromagnetic field}
\DeclareAcronym{EM}{short = EM , long = electromagnetic}
\DeclareAcronym{ETSI}{short = ETSI , long = European Telecommunications Standards Institute}
\DeclareAcronym{E-field}{short = E-field , long = electric field}

\DeclareAcronym{FCC}{short = FCC , long = Federal Communications Commission}
\DeclareAcronym{FDD}{short = FDD , long = frequency division duplex}
\DeclareAcronym{FR1}{short = FR1 , long = frequency range 1}
\DeclareAcronym{FR2}{short = FR2 , long = frequency range 2}
\DeclareAcronym{FSPL}{short = FSPL , long = free-space path loss}

\DeclareAcronym{GO}{short = GO , long = geometrical optics}
\DeclareAcronym{GPU}{short = GPU , long = graphics processor unit}
\DeclareAcronym{GAS}{short = GAS , long = geometry acceleration structure}
\DeclareAcronym{GA}{short = GA , long = genetic algorithm}

\DeclareAcronym{HDII}{short = HDII , long = half-distance insersion and interpolation}
\DeclareAcronym{ICES}{short = ICES , long = International Committee on Electromagnetic Safety}
\DeclareAcronym{ICNIRP}{short = ICNIRP , long = International Commission on Non-Ionizing Radiation Protection}
\DeclareAcronym{IEC}{short = IEC , long = International Electrotechnical Commission}
\DeclareAcronym{IEEE}{short = IEEE , long = Institute of Electrical and Electronics Engineers}
\DeclareAcronym{IMT}{short = IMT , long = International Mobile Telecommunications}
\DeclareAcronym{IoT}{short = IoT , long = Internet of Things}
\DeclareAcronym{ITU}{short = ITU , long = International Telecommunication Union}
\DeclareAcronym{IHE}{short = IHE , long = Institute of Radio Frequency Engineering and Electronics}
\DeclareAcronym{IS}{short = IS , long = intersection}

\DeclareAcronym{KIT}{short = KIT , long = Karlsruhe Institute of Technology}

\DeclareAcronym{LoS}{short = LoS , long = line-of-sight}
\DeclareAcronym{LTE}{short = LTE , long = long term evolution}

\DeclareAcronym{mMIMO}{short = mMIMO , long = massive multiple input multiple output}
\DeclareAcronym{mmWave}{short = mmWave , long = millimeter wave}
\DeclareAcronym{mRRH}{short = mRRH , long = micro-RRH}
\DeclareAcronym{MSS}{short = MSS , long = mobile satellite service}
\DeclareAcronym{MIMO}{short = MIMO , long = multiple input multiple output}
\DeclareAcronym{MADS}{short = MADS , long = mesh adaptive direct search}
\DeclareAcronym{MRT}{short = MRT , long =  maximum ratio transmission}

\DeclareAcronym{NCRP}{short = NCRP , long = National Council on Radiation Protection and Measurements}
\DeclareAcronym{NG-RAN}{short = NG-RAN , long = next generation radio access network}
\DeclareAcronym{NR}{short = NR , long = new radio}
\DeclareAcronym{NSA}{short = NSA , long = non-stand-alone}
\DeclareAcronym{NTN}{short = NTN , long = non-terrestrial network}
\DeclareAcronym{NM}{short = NM , long = Nelder-Mead}
\DeclareAcronym{NLoS}{short = NLoS , long = non line-of-sight}

\DeclareAcronym{OSM}{short = OSM , long = OpenStreetMap}

\DeclareAcronym{PBCH}{short = PBCH , long = physical broadcast channel}
\DeclareAcronym{PCI}{short = PCI , long = physical cell ID}
\DeclareAcronym{PDSCH}{short = PDSCH , long = physical downlink shared channel}
\DeclareAcronym{pRRH}{short = pRRH , long = pico-RRH}
\DeclareAcronym{PTX}{short = PTX , long = parallel thread execution}
\DeclareAcronym{PDP}{short = PDP , long = power delay profile}
\DeclareAcronym{PLE}{short = PLE , long = path loss exponent}

\DeclareAcronym{QoE}{short = QoE , long =  quality of experience }
\DeclareAcronym{QoS}{short = QoS , long =  quality of service }

\DeclareAcronym{RAN}{short = RAN , long = radio access network}
\DeclareAcronym{RAT}{short = RAT , long = radio access technology}
\DeclareAcronym{RF}{short = RF , long = radio frequency}
\DeclareAcronym{RRH}{short = RRH , long = remote radio heads}
\DeclareAcronym{RSS}{short = RSS , long = root sum square}
\DeclareAcronym{RL}{short = RL , long = ray-launching}
\DeclareAcronym{RT}{short = RT , long = ray-tracing}
\DeclareAcronym{Rx}{short = Rx , long = receiver}
\DeclareAcronym{RMSE}{short = RMSE , long = root mean squared error}
\DeclareAcronym{RSSI}{short = RSSI , long = received signal strength indicator}
\DeclareAcronym{RSRP}{short = RSRP , long = reference signal received power}

\DeclareAcronym{SA}{short = SA , long = stand-alone}
\DeclareAcronym{SAN}{short = SAN ,  long = satellite access node}
\DeclareAcronym{SAR}{short = SAR , long = specific absorption rate}
\DeclareAcronym{SSB}{short = SSB , long = synchronization signal block}
\DeclareAcronym{SBR}{short = SBR , long = shoot and bouncing ray}
\DeclareAcronym{SBT}{short = SBT , long = shader binding table}
\DeclareAcronym{std}{short = std , long = standard deviation}
\DeclareAcronym{SIR}{short = SIR , long = signal-to-interference ratio}
\DeclareAcronym{SNR}{short = SNR , long = signal-to-noise ratio}

\DeclareAcronym{TDD}{short = TDD , long = time division duplex}
\DeclareAcronym{THz}{short = THz , long = terahertz}
\DeclareAcronym{TN}{short = TN , long = terrestrial network}
\DeclareAcronym{TUK}{short = TUK, long = Technische Universit\"at Kaiserslautern}
\DeclareAcronym{Tx}{short = Tx , long = transmitter}

\DeclareAcronym{UAS}{short = UAS , long = unmanned aerial system }
\DeclareAcronym{UE}{short = UE , long = user equipment}
\DeclareAcronym{UTD}{short = UTD , long = uniform theory of diffraction}

\DeclareAcronym{VSAT}{short = VSAT , long = very small aperture terminal }
\DeclareAcronym{VPL}{short = VPL , long = vertical-plane-launch}

\DeclareAcronym{WHO}{short = WHO , long = World Health Organization}

\DeclareAcronym{ZF}{short = ZF , long = zero-forcing}

\usepackage{fancyhdr}
\fancypagestyle{empty}{fancy}


\begin{document}

\title{Multi-RIS Deployment Optimization for mmWave ISAC Systems in Real-World Environments}
\author{Yueheng Li,~\IEEEmembership{Member,~IEEE,} Xueyun Long,~\IEEEmembership{Graduate Student Member,~IEEE,} Mario Pauli,~\IEEEmembership{Senior Member,~IEEE,}  Suheng Tian,~\IEEEmembership{Student Member,~IEEE,} Xiang Wan,~\IEEEmembership{Senior Member,~IEEE,} Benjamin Nuss,~\IEEEmembership{Senior Member,~IEEE,} Tiejun Cui,~\IEEEmembership{Fellow,~IEEE,} Haixia Zhang,~\IEEEmembership{Senior Member,~IEEE,} and Thomas Zwick,~\IEEEmembership{Fellow,~IEEE}
\vspace{-1.7\baselineskip}}


\maketitle
\begin{abstract}

Reconfigurable intelligent surface-assisted integrated sensing and communication (RIS-ISAC) presents a promising system architecture to leverage the wide bandwidth available at millimeter-wave (mmWave) frequencies, while mitigating severe signal propagation losses and reducing infrastructure costs. To enhance ISAC functionalities in the future air-ground integrated network applications, RIS deployment must be carefully designed and evaluated, which forms the core motivation of this paper. To ensure practical relevance, a multi-RIS-ISAC system is established, with its signal model at mmWave frequencies demonstrated using ray-launching calibrated to real-world environments. On this basis, an energy-efficiency-driven optimization problem is formulated to minimize the multi-RIS size-to-coverage sum ratio, comprehensively considering real-world RIS deployment constraints, positions, orientations, as well as ISAC beamforming strategies at both the base station and the RISs. To solve the resulting non-convex mixed-integer problem, a simplified reformulation based on equivalent gain scaling method is introduced. A two-step iterative algorithm is then proposed, in which the deployment parameters are determined under fixed RIS positions in the first step, and the RIS position set is updated in the second step to progressively approach the optimum solution. Simulation results based on realistic parameter benchmarks present that the optimized RISs deployment significantly enhances communication coverage and sensing accuracy with the minimum RIS sizes, outperforming existing approaches.

\end{abstract}

\section{Introduction}
\label{introduction}
The next-generation mobile communication technologies aim to establish air-ground integrated networks to support a wide range of applications, including vehicle-to-everything, the internet of things, etc. To meet such demands, integrated sensing and communication (ISAC) has emerged as an essential technology, leveraging the advantage of combining dual functionalities into a unified system \cite{de2021joint}, thereby enabling efficient collaborations and reducing infrastructure costs. In this context, to fulfill the requirements for higher communication data rates and sensing accuracy, the desire of wider bandwidths drives 3GPP standardization at millimeter-wave (mmWave) frequencies \cite{giordani2018tutorial}. However, the wireless signal propagation at mmWave frequencies is significantly hindered by free-space path loss (FSPL), reflection loss, and diffraction loss, all of which constrain effective signal coverage and strength. Typical compensation solutions, such as the dense deployment of base stations (BSs) or increased transmit power, yield high-cost drawback, which is even more pronounced in the expected large-scale, multi-dimensional air-ground integrated networks.

In recent years, reconfigurable intelligent surface (RIS) has attracted significant research attentions. A RIS constructed with low-cost micro-components-based unit cells in a uniform planar array (UPA) geometry, can be conveniently deployed to enable controllable reflective beamforming. This technology could reconstruct the wireless channel to enhance signal coverage and strength within a favorable budget, offering a promising solution to the aforementioned cost-performance trade-off challenge faced by mmWave ISAC systems, particularly in urban environments with substantial blockages. The concept of RIS was first introduced in \cite{cui2014coding}, which has been further developed to serve various applications, such as reflective relay \cite{wu2021intelligent}, multi-path enhancement \cite{alexandropoulos2023ris}, etc. In this trend, fruitful research contributions in RIS technologies have emerged, including RIS antenna designs \cite{wan2021reconfigurable,9889185}, beamforming strategies for RIS-assisted communication and RIS-assisted ISAC (RIS-ISAC) systems \cite{ma2022cooperative,luo2022joint}, as well as prototype system demonstrations that validate the feasibility of RIS-ISAC signal propagation \cite{9999288,10438390}.

While existing studies have strongly presented the potential of RIS to serve future wireless communication networks, the deployment of RIS naturally becomes a key research focus when approaching practical implementations. It is widely recognized that RISs are typically deployed on building side walls, or carried by unmanned aerial vehicles (UAVs) \cite{liu2023integrated}. The latter option benefits from the high spatial freedom of UAV flight, making trajectory design a prominent research direction \cite{ wu2024joint}. However, the building-attached option is inherently less flexible, as once installed, the RIS panels cannot be easily reconstructed, which necessitates careful design and thorough evaluation before deployment. In representative RIS deployment studies, \cite{10172310} determined the RIS position based on the minimum path loss criterion between the BS and user equipment (UE) along a 1D line, while \cite{9687840} extended the study to RIS placement in a 2D area maximizing the achievable rate, and \cite{yu2023active} considered the sensing performance following the similar logic. Upon this, RIS deployment was optimized for non-orthogonal multiple access networks in \cite{liu2020ris}, contributing to more concrete applications. Besides deployment position, the RIS orientation optimization to enhance communication signal coverage was explored in \cite{zeng2020reconfigurable}. To address urban scenarios with significant blockages, the optimization of multiple RIS deployments was examined in \cite{ma2024multi}. Rather than statistical scenarios, the authors jointly considered the aforementioned deployment parameters in a real-world environment in \cite{fu2025multi}. Besides communication orientated studies, in the most recent and concurrent study with this paper, \cite{encinas2025isac} started the RIS deployment study serving ISAC functionalities with the path loss criterion inspired by \cite{10172310, 9687840, yu2023active}.

Through the review of the state-of-the-art, further RIS deployment studies to serve future ISAC system could be applied in the following aspects. First, the quality of the received signal via RIS assistance is intricately dependent on the attenuation factors of the cascaded channel, the signal incidence and reflection angles to/from the RIS, the RIS unit cell radiation properties, and the BS+RIS beamforming strategy \cite{9837936}. Therefore, upon existing studies focusing on isolated features, RIS deployment optimization should be approached through a more comprehensive model that incorporates environment, spatial, and signal processing features to multiple RISs in high-fidelity scenarios at mmWave frequencies. Second, to meet the energy efficiency demand and further reduce the infrastructure costs of future wireless communication networks, a more refined optimization of the deployed number of RIS unit cells is critical, ensuring that RIS assists ISAC in the most efficient manner. Last but not least, while both RIS and ISAC are essential technologies, studies on RIS deployment specifically tailored for ISAC functionalities remain limited. Further research is urgently needed to deepen the understanding of the quantitative relationship between RIS deployment and ISAC performance, to explore effective optimization strategies, and to clarify the necessity as well as impact of RIS deployment in practical scenarios.

To realize the aforementioned ambitions, in this paper, the optimization of multi-RIS deployment for mmWave ISAC system in real-world environments is proposed. The main contributions of this paper are listed as follows:

\begin{itemize}
    \item A RIS-ISAC system is established for an air–ground integrated network in real-world environments. With ray-launching-based channel modeling, the ISAC signal model at mmWave frequencies is formulated, validating both communication coverage analysis for UEs and sensing accuracy evaluation for UAV.
    \item An energy-efficiency-driven optimization problem is formulated to minimize the multi-RIS size-to-coverage sum ratio, equivalently representing the minimum number of RIS unit cells deployed for the serving area, while ensuring the required communication coverage and sensing accuracy. For comprehensive study, this formulation jointly consider RIS geometric properties—including real-world placement constraints, deployment positions, and orientations—as well as signal processing aspects including BS beamforming and RIS beamforming for ISAC functionalities.
    \item A method is proposed to address the aforementioned non-convex, mixed-integer, real-world mapping problem. It first simplifies the original formulation using an equivalent gain scaling method, and then solves the reformulated problem through two-step iterative algorithm: The first step determines the optimization variables for RIS at given position set, and the second step updates the position set based on the obtained configurations to progressively approach the optimal solution.
    \item Simulations with realistic parameters demonstrate that the proposed RIS deployment optimization significantly enhances communication signal coverage while enabling accurate sensing. In comparison to existing approaches, the proposed strategy incorporates a more complete and practical set of RIS deployment properties, achieving full dual-functionality with reduced RIS sizes, highlighting the necessity of RIS-ISAC deployment studies.
\end{itemize}

The remainder of this paper is organized as follows. Section~\ref{system} introduces the air–ground integrated RIS-ISAC system model. Section~\ref{form} formulates the RIS deployment optimization problem, and Section~\ref{proposed_algo} presents the proposed solution. Section~\ref{sim} provides the simulation results, and Section~\ref{sec:conc} concludes the paper and outlines directions for future work.

For mathematical notation: $\otimes$ is the Kronecker product. $(\cdot)^{*}$, $(\cdot)^{T}$, $(\cdot)^{H}$ stand for complex conjugate, transpose, and Hermitian transpose operations, respectively. $\angle[\cdot]$, $||\cdot||$, $\mathbb{E}[\cdot]$, $\text{Tr}(\cdot)$, $\Re\{\cdot\}$ denote the phase of a complex number, the Euclidean norm, the expectation, the trace and real-part operator. 

\section{System Model}
\label{system}
In this section, the proposed RIS-ISAC system in real-world environments is presented. Upon this, the communication and sensing signal models are introduced, laying the foundation for RIS deployment optimization.

\vspace{-0.5\baselineskip}
\subsection{RIS-ISAC Application Scenario}\label{tool}
\begin{figure*}[!t]
 \vspace*{-8pt} 
\centering
\subfigure[]{\includegraphics[width=3.4in, trim=3mm 2mm 10mm 3mm,clip ]{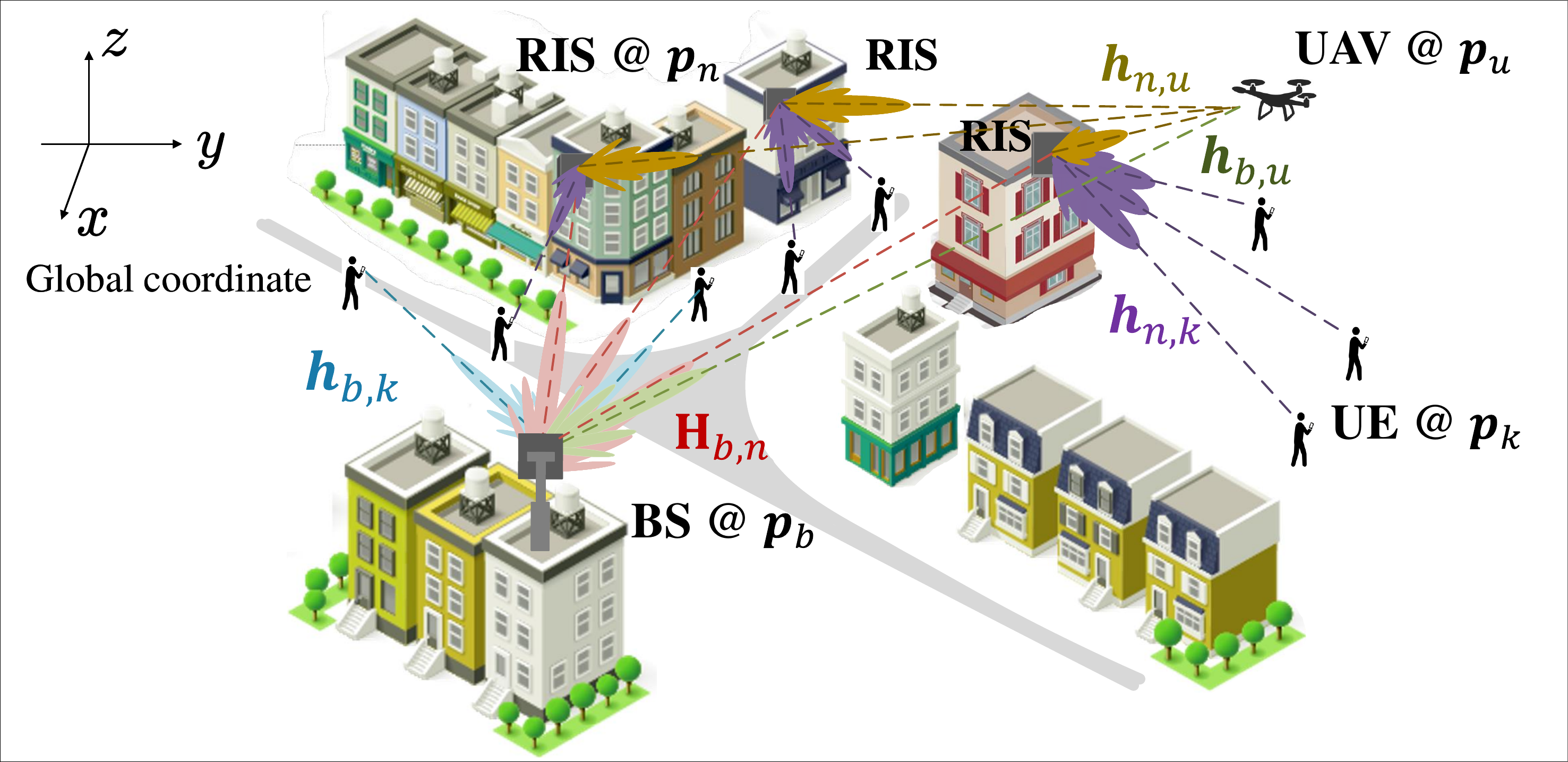}%
\label{isac_app_scen}
}
\hspace{-15pt}
\subfigure[]{\includegraphics[width=1.9in, trim=20mm 60mm 5mm 3mm,clip ]{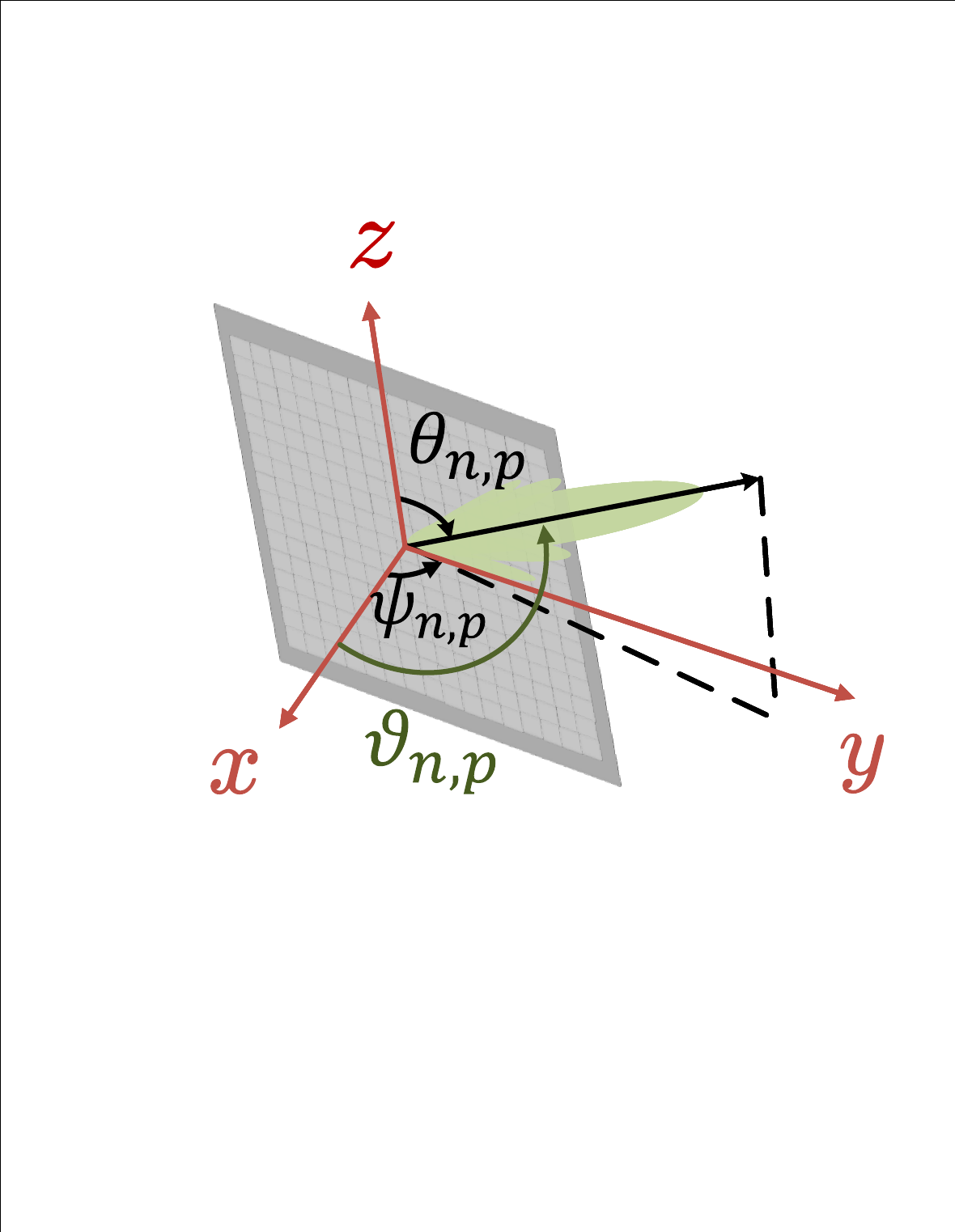}%
\label{UPA_loc_coord}
}
\hspace{-15pt}
\subfigure[]{\includegraphics[width=1.9in, trim=3mm 60mm 10mm 3mm,clip]{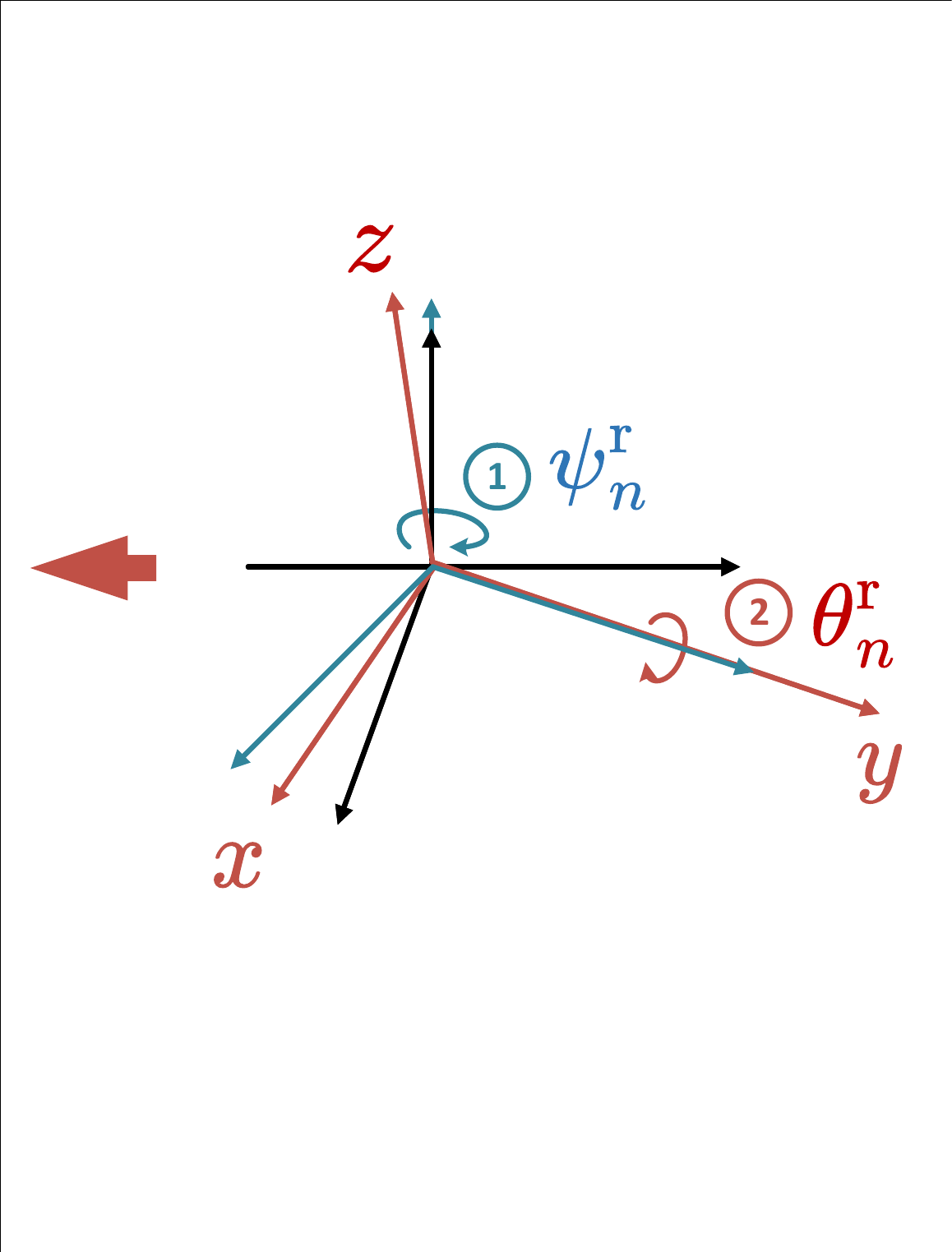}%
\label{RIS_ori_setting}
}
\caption{(a) The proposed RIS-ISAC system model in an urban environment. The positions of the network nodes are defined in global coordinates, while each UPA has local coordinates in the $yz$-plane, with its facing direction (normal) aligned with the $x$-axis.
(b) UPA local coordinates with angle definitions using RIS as an example. The term $\vartheta_{n,p}$ represents the angle between the $n$-th RIS DoD/DoA ($\psi_{n,p}$ in the azimuth dimension and $\psi_{n,p}$ in elevation dimension of the $xy$-plane) to node $p \in \{b,k,u\}$ and the RIS normal.(c) RIS orientation definition. The global coordinate system (\textcolor{black}{$\rightarrow$}) first rotates around the $z$-axis by $\psi_n^{\text{r}}$ (\textcolor{cyan!70!black}{$\rightarrow$}) then rotates around the $y$-axis by $\theta_n^{\text{r}}$ (\textcolor{red!70!black}{$\rightarrow$}), which ends up with the local coordinate as in (b).}
\label{scenario}
\end{figure*}

The proposed air-ground integrated RIS-ISAC system model is illustrated in Fig.~\ref{isac_app_scen}. A BS located at $\boldsymbol{p}_b = [x_b,y_b,z_b]$ attempts to communicate with $K$ UEs at $\boldsymbol{p}_k = [x_k,y_k,z_k], k \in \{1,2,...,K\}$ in the ground layer. Simultaneously, the BS needs to sense the location and velocity of an UAV at $\boldsymbol{p}_u = [x_u,y_u,z_u]$ in the air layer. Due to signal blockages in dense urban environments and the severe FSPL, reflection loss, and diffraction loss at mmWave frequencies, RISs are deployed to enhance communication signal coverage and improve sensing accuracy (written as ISAC QoS for simplification in later parts of the paper) \cite{wu2021intelligent}.

As future wireless communication networks aim to ubiquitous connectivity, energy efficiency becomes a critical consideration in system design. Given that practical RISs often involve large panels \cite{xu2020resource}, optimizing RIS deployment to meet the required ISAC quality of service (QoS) while minimizing the utilized number of unit cells is a sustainable and effective strategy. To achieve this goal under practical application scenarios, this paper utilizes the deterministic ray-launching tool, namely IHE-RL, developed at the Karlsruhe Institute of Technology in our previous study \cite{long2025joint}. Unlike the statistical channel models in \cite{10172310,ma2024multi,9837936}, which serve as efficient starting points for RIS deployment studies, IHE-RL accurately derives wireless channel coefficients based on 3D urban environments from OpenStreetMap (OSM), including line-of-sight (LoS) transmission, reflection, diffraction, and the consideration of dielectric loss, enhancing adaptability for real-world environments. The accuracy of IHE-RL was validated through extensive simulations and measurements in \cite{long2025joint}, which can be referred for interests.

\vspace{-0.5\baselineskip}
\subsection{Communication Signal Model}\label{System Model}
According to the path-based multi-antenna channel model in \cite{eisenbeis2018path}, the communication channel $\boldsymbol{h}_{\text{C},b,k} \in \mathbb{C}^{1 \times M_b}$ from the BS with an UPA to the $k$-th UE (BS2UE) with a single antenna can be expressed as
\begin{align}\label{BSmultipathchannelsingleUE}
    \boldsymbol{h}_{\text{C},b,k} = \sum_{n_p=0}^{N_p-1} &F_k(\theta_{k,n_p},\psi_{k,n_p}) a_k(\theta_{k,n_p},\psi_{k,n_p}) \alpha_{n_p} e^{j \varphi_{n_p}} \cdot \notag\\
    & \boldsymbol{a}_b^T(\theta_{b,n_p},\psi_{b,n_p}) \textbf{F}_b(\theta_{b,n_p},\psi_{b,n_p}) \notag\\
    = \sum_{n_p=0}^{N_p-1} &G_k(\theta_{k,n_p},\psi_{k,n_p}) a_k(\theta_{k,n_p},\psi_{k,n_p}) \alpha_{n_p} e^{j \varphi_{n_p}} \cdot \notag\\
    & \boldsymbol{a}_b^T(\theta_{b,n_p},\psi_{b,n_p})G_b(\theta_{b,n_p},\psi_{b,n_p}),
\end{align}
where $N_p$ is the total number of existing paths indexed by $n_p$, associated with an attenuation factor $\alpha_{n_p}$ and a phase $\varphi_{n_p}$. The scalar $F_k(\theta_{k,n_p},\psi_{k,n_p})$ represents the single antenna radiation pattern at the UE in the direction of arrival (DoA) $(\theta_{k,n_p},\psi_{k,n_p})$ as depicted in Fig.~\ref{UPA_loc_coord}. Similarly, the diagonal matrix $\textbf{F}_b(\theta_{b,n_p},\psi_{b,n_p}) \in\mathbb{R}^{M_b\times M_b}$ represents the multi-antenna radiation pattern contributions in the direction of departure (DoD) $(\theta_{b,n_p},\psi_{b,n_p})$, based on the UPA local coordinate in Fig.~\ref{UPA_loc_coord}. With normalized efficiency, the radiation patterns can be denoted by the receive (Rx) and transmit (Tx) antenna gains $G_k(\theta_{k,n_p},\psi_{k,n_p})$ and $G_b(\theta_{b,n_p},\psi_{b,n_p})$ in the second step of \eqref{BSmultipathchannelsingleUE}. The BS UPA contains $M_b = M_b^{\text{y}}\times M_b^{\text{z}}\geq K$ antenna elements in the $yz$-plane will lead to the steering vector $\boldsymbol{a}_{b}(\theta_{b,n_p},\psi_{b,n_p})\in \mathbb{C}^{M_b \times 1}$ in \eqref{BSmultipathchannelsingleUE} expressed as
\begin{equation} \label{DoD}
\boldsymbol{a}_{b}(\theta_{b,n_p}, \psi_{b,n_p}) = \boldsymbol{a}_y(\theta_{b,n_p}, \psi_{b,n_p}) \otimes \boldsymbol{a}_z(\theta_{b,n_p}, \psi_{b,n_p}),
\end{equation}
with
\begin{align}  \label{DoD_y}
    \boldsymbol{a}_y(\theta_{b,n_p}, \psi_{b,n_p}) = 
    [&1,\dots,e^{j\kappa D (m_b^{\text{y}}-1) \sin\theta_{b,n_p} \sin\psi_{b,n_p}}, \notag\\
    &\dots, e^{j\kappa D (M_b^{\text{y}}-1) \sin\theta_{b,n_p} \sin\psi_{b,n_p}}]^T,
\end{align}
and
\begin{align} \label{DoD_z}
\boldsymbol{a}_z(\theta_{b,n_p}, \psi_{b,n_p}) = [&1, \dots,e^{j\kappa D (m_b^{\text{z}}-1) \cos\theta_{b,n_p}}, \notag\\
&\dots,e^{j\kappa D (M_b^{\text{z}}-1) \cos\theta_{b,n_p}}]^T,
\end{align}
where $m_b^{\text{y}}\in\{1,2,...,M_b^{\text{y}}\}$ and $m_b^{\text{z}}\in\{1,2,...,M_b^{\text{z}}\}$.
The term $\kappa = 2\pi /\lambda_c$ is the wave number, and $D=\lambda_c/2$ denotes the UPA element spacing based on the wavelength $\lambda_c = c_0/f_c$ at the center carrier frequency $f_c$ with light speed $c_0$. For single antenna UE, $a_k(\theta_{k,n_p}, \psi_{k,n_p}) = 1$. 

Extending \eqref{BSmultipathchannelsingleUE} to the communication channel matrix between the BS and $K$ UEs $\textbf{H}_{\text{C}}=[\boldsymbol{h}_{\text{C},b,1}^{T},...,\boldsymbol{h}_{\text{C},b,k}^{T},...,\boldsymbol{h}_{\text{C},b,K}^{T}]^{T}$, the received signal $\boldsymbol{y}_{\text{C}} = [y_{C,1},...,y_{C,k},...,y_{C,K}]^T$ at the UEs via the BS2UE channel is
\begin{equation}\label{RxsignalonlyBS}
\boldsymbol{y}_{\text{C}}=\textbf{H}_\text{C}\textbf{W}_\text{C}\boldsymbol{s}_{\text{C}} + \boldsymbol{n},
\end{equation}
where $\textbf{W}_{\text{C}} = [\boldsymbol{w}_{b,1},\dots, \boldsymbol{w}_{b,k},...,\boldsymbol{w}_{b,K}] \in \mathbb{C}^{M_b \times K}$ denotes the communication beamforming matrix at the BS, with $\boldsymbol{w}_{b,k} \in\mathbb{C}^{M_b\times 1}$ being the beamforming vector for the $k$-th UE. The vector $\boldsymbol{s}_{\text{C}} = [s_1,...,s_k,...,s_K]^T \in \mathbb{C}^{K \times 1}$ denotes the normalized transmit signal, and $\boldsymbol{n} = [n_1,...,n_k,...,n_K]^T \in \mathbb{C}^{K \times 1}$ represents the noise term having $n_k \sim \mathcal{CN}(0, \sigma_{n_k}^2)$, with $\sigma_{n_k}^2$ being the noise power spectral density at the $k$-th UE. Therefore, the signal-to-noise ratio (SNR) of the $k$-th UE is expressed as
\begin{equation}\label{Only BS SINR}
    \gamma_{\text{C},b,k} = \frac{P_k|\boldsymbol{h}_{\text{C},b,k}\boldsymbol{w}_{b,k}|^2}{\sigma_{n_k}^2},
\end{equation}
where $P_{k}$ is the BS transmit power allocated for the $k$-th UE. 

For future ISAC systems operating at mmWave frequencies in urban environments, even with a properly designed $\boldsymbol{w}_{b,k}$, the UE SNR may still fall below a lower bound $\gamma^{\text{tr}}$ even for interference-free case due to the high probability of LoS blockage caused by obstacles, coupled with severe propagation losses mentioned in \ref{tool}. Therefore, placing RISs to create dominant cascaded paths from BS to RISs (BS2RIS) and from RIS to UEs (RIS2UE) can effectively enhance the signal coverage. In this context, when the communication between the BS and the $k$-th UE is assisted by the $n$-th RIS phase centered at $\boldsymbol{p}_{n} = [x_{n}, y_{n}, z_{n}]$ among $N$ total RISs, the received signal is reformed as
\begin{equation}\label{RxsignalwithRIS}
y_{\text{C},b,n,k} = \boldsymbol{h}_{n,k}\mathbf{\Phi}_{n}\textbf{H}_{b,n}\boldsymbol{w}_{b,n} s_k + n_k.
\end{equation}
In an air–ground integrated ISAC scenario, where horizontal and vertical beamforming are equally important, the RIS is modeled as a square UPA with $M_{n,s}$ unit cells along each axis, resulting in $M_n = M_{n,s}^2$ unit cells spaced by $D$. Considering the far-field signal propagation and the utilization of practical low-cost RIS without amplitude control \cite{wan2021reconfigurable}, the RIS beamforming matrix is defined as $\boldsymbol{\Phi}_n \in \mathbb{C}^{M_n \times M_n} = \text{diag}\{\boldsymbol{\phi}_n\}$, where $\boldsymbol{\phi}_n = [e^{j\epsilon_{n,1}}, ..., e^{j\epsilon_{n,m_n}}, ..., e^{j\epsilon_{n,M_n}}]^T$ with phase shift $\epsilon_{n,m_n}$ applied at the $m_n$-th RIS unit cell. Given that RIS employs analog beamforming, the unit cells have $L$-bit phase resolution, yielding $\epsilon_{n,m_n} \in \{{\frac{2\pi l}{2^L} | l = 0, 1, ..., 2^L-1}\}$ based on an uniform phase quantization. Back to \eqref{RxsignalwithRIS}, $\boldsymbol{w}_{b,n} \in \mathbb{C}^{M_b \times 1}$ denotes the BS2RIS beamforming vector, and $\boldsymbol{h}_{n,k} \in\mathbb{C}^{1\times M_n}$ denotes the RIS2UE channel, written in details as
\begin{align}\label{RIS2UE}
\boldsymbol{h}_{n,k} &= G_k(\theta_{k,n},\psi_{k,n}) \alpha_{n,k} e^{j \varphi_{n,k}} \cdot \notag\\
&\boldsymbol{a}_n^T(\theta_{n,k},\psi_{n,k}) G_n(\theta_{n,k},\psi_{n,k}).
\end{align}
Here, the UE antenna gain $G_k(\theta_{k,n},\psi_{k,n})$, the RIS reflection gain $G_n(\theta_{n,k},\psi_{n,k})$, the steering vector $\boldsymbol{a}_n(\theta_{n,k},\psi_{n,k})$, the attenuation factor $ \alpha_{n,k}$ and phase $\varphi_{n,k}$ follow a similar derivation to those in \eqref{BSmultipathchannelsingleUE}, by defining the Tx and Rx to be RIS and UE. Further for \eqref{RxsignalwithRIS}, the BS2RIS channel matrix is denoted by
\begin{align}\label{BS2RIS}
\textbf{H}_{b,n} = &G_n(\theta_{n,b},\psi_{n,b})\boldsymbol{a}_n(\theta_{n,b},\psi_{n,b}) \alpha_{b,n} e^{j \varphi_{b,n}} \cdot \notag\\
&\boldsymbol{a}_b^T(\theta_{b,n},\psi_{b,n}) G_b(\theta_{b,n},\psi_{b,n}).
\end{align}
Again, by defining Tx and Rx of \eqref{BSmultipathchannelsingleUE} as BS and RIS, the RIS incident gain $G_n(\theta_{n,b},\psi_{n,b})$, the BS antenna gain $G_b(\theta_{b,n},\psi_{b,n})$, the steering vectors $\boldsymbol{a}_n(\theta_{n,b},\psi_{n,b})$ and $\boldsymbol{a}_b(\theta_{b,n},\psi_{b,n})$, the attenuation factor $\alpha_{b,n}$ and phase $\varphi_{b,n}$ can be derived. Please notice that \eqref{RIS2UE}  and \eqref{BS2RIS} have no summation product as \eqref{BSmultipathchannelsingleUE} since BS2RIS and RIS2UE has dominate path for proper RIS deployment \cite{wu2021intelligent} (Also see Section \ref{RISposition}).
Now, the SNR of the $k$-th UE with the assistance of the $n$-th RIS becomes
\begin{equation}\label{RIS_SINR}
    \gamma_{\text{C},b,n,k} = \frac{P_{n}|\boldsymbol{h}_{\text{C},b,n,k}\boldsymbol{w}_{b,n}|^2}{\sigma_{n_k}^2},
\end{equation}
with $P_{n}$ being the BS transmit power assigned to the cascade channel
\begin{equation}\label{caschannel}
\boldsymbol{h}_{\text{C},b,n,k}= \boldsymbol{h}_{n,k}\boldsymbol{\Phi}_{n}\textbf{H}_{b,n}.
\end{equation}

\vspace{-0.5\baselineskip}
\subsection{Sensing Signal Model}\label{Sensing Model}
Although mmWave ISAC systems offer wide bandwidth for enhanced sensing resolution, significant losses degrade the sensing SNR, yielding reduced accuracy. To mitigate this, RISs can introduce additional reflection paths that coherently combine sensing signals, assuming multi-path delays are negligible relative to the radar symbol duration \cite{10243495}. However, this assumption constrains sensing resolution, rendering it inadequate for future applications. Therefore, this paper proposes a more practical strategy that applies sensing performance boundary for every RIS-assisted path, leveraging known RIS positions to validate multi-path sensing, as outlined below.

For mono-static sensing, the received signal expressed in delay-Doppler format \cite{10438390} can be written as
\begin{align}\label{SensingsignalwithRIS}
    y_{\text{S}}(t)= \sum_{n = 0}^{N} \alpha_n e^{j2\pi f_{D_n}t}s_{k}(t-\tau_{n}) + n_k(t).
\end{align}
Considering that every RIS in the environment would contribute to the UAV sensing as in Fig.~\ref{isac_app_scen}, there exists in total $N+1$ dominating paths, with $n=0$ denoting the direct sensing path from BS to UAV (BS2UAV), and $n \in \{1,2,...,N\}$ corresponding to the each RIS-assisted sensing path. The term $\tau_{n}$ denotes the round-trip propagation delay, and $f_{D_{n}}$ is the corresponding Doppler shift induced by the UAV movement. By implementing Fourier Transform-based sensing signal processing \cite{brunner2024bistatic}, $\tau_n$ and $f_{D_n}$ will be presented as range information $d_{n} = c_0\tau_n/2$ and velocity information $v_{n} = \lambda_c f_{D_n}/2$ . The term $\alpha_n$ in \eqref{SensingsignalwithRIS} is the effective channel coefficient, separately categorized with BS2UAV and RIS-assisted path as
\begin{equation} \label{def_alpha_n}
    \alpha_n \!= \!\left\{
    \begin{aligned}
        &\!\alpha_u \boldsymbol{w}_{b,u}^T \textbf{H}_{\text{S},b,u}\boldsymbol{w}_{b,u}, \quad n= 0\\
        &\!\alpha_u \boldsymbol{w}_{b,u}^{T}\textbf{H}_{\text{S},b,n,u}\boldsymbol{w}_{b,n} \!+\! \alpha_u \boldsymbol{w}_{b,n}^{T}{\textbf{H}^{T}_{\text{S},b,n,u}}\!\boldsymbol{w}_{b,u},n \neq 0.
    \end{aligned}
    \right. 
\end{equation}
Here, $\alpha_u$ represents the radar cross section (RCS) of the UAV, and $\boldsymbol{w}_{b,u} \in \mathbb{C}^{M_b \times 1}$ is the BS2UAV beamforming vector. For UAV in the air-layer,  the LoS path dominates the BS2UAV effective channel
\begin{equation}\label{BStoUAVsensing}
    \textbf{H}_{\text{S},b,u} =  \boldsymbol{h}_{b,u}^T \boldsymbol{h}_{b,u},
\end{equation}
considering sensing channel channel reciprocity, where
\begin{align}\label{BS2UAVchannel}
    \boldsymbol{h}_{b,u} =\alpha_{b,u} e^{j \varphi_{b,u}} \boldsymbol{a}_b^T(\theta_{b,u},\psi_{b,u}) G_b(\theta_{b,u},\psi_{b,u}),
\end{align}
denoting the incident channel. Following the definitions in \eqref{BSmultipathchannelsingleUE}, the term $G_b(\theta_{b,u},\psi_{b,u})$ is the equivalent antenna gain along DoD $(\theta_{b,u},\psi_{b,u})$, $\alpha_{b,u}$ and $\varphi_{b,u}$ are the attenuation factor and phase. For each RIS-assisted effective channel coefficient in \eqref{def_alpha_n}, it is the sum of two cascaded paths with identical delay and Doppler information, yielding the channel coefficients of the BS $\rightarrow$ RIS $\rightarrow$ UAV $\rightarrow$ BS path
\begin{equation}\label{SensingchannelwithRIS}
    \textbf{H}_{\text{S},b,n,u} = \boldsymbol{h}_{b,u}^T \boldsymbol{h}_{n,u} \boldsymbol{\Phi}_{n} \textbf{H}_{b,n},
\end{equation}
and the BS $\rightarrow$ UAV $\rightarrow$ RIS $\rightarrow$ BS path ${\textbf{H}^{T}_{\text{S},b,n,u}}$ considering sensing channel reciprocity. The vector $\boldsymbol{h}_{n,u} \in \mathbb{C}^{1 \times M_n}$ indicates the RIS2UAV channel
\begin{equation} \label{RIS2UAV}
    \boldsymbol{h}_{n,u} = \alpha_{n,u} e^{j \varphi_{n,u}} \boldsymbol{a}_n^T(\theta_{n,u},\psi_{n,u}) G_n(\theta_{n,u},\psi_{n,u}).
\end{equation}
Similar to the coefficients derivations in \eqref{BSmultipathchannelsingleUE}, $(\theta_{n,u},\psi_{n,u})$, $\alpha_{n,u}$ and $\varphi_{n,u}$ denote the DoD from the RIS, attenuation factor and phase, respectively. It is worth to notice that although the  BS $\rightarrow$ RIS $\rightarrow$ UAV $\rightarrow$ RIS $\rightarrow$ BS path exists, it has significantly stronger attenuation due to the additional cascaded path loss verified by the IHE-RL tool, therefore neglected in \eqref{def_alpha_n}.

For signal processing, OFDM is adopted as the ISAC waveform in this paper with the number of subcarriers $N_c$, the number of OFDM symbols $M_{\text{OFDM}}$, and the bandwidth $B$. Since cyclic prefix (CP) does not provide radar processing gain, it is omitted here assuming sufficient length applied. Therefore, the effective duration of a single OFDM symbol is $T_{\text{OFDM}} = N_c/B$, yielding the total frame duration $T_0 = M_{\text{OFDM}} T_{\text{OFDM}}$.
To evaluate the sensing accuracy for range and velocity, the parameter to be estimated is $\boldsymbol{\eta} = [\boldsymbol{\eta}_0, ..., \boldsymbol{\eta}_n, ..., \boldsymbol{\eta}_N]^T$, where $\boldsymbol{\eta}_n = [d_n,v_n]^T$. With every path having a distinct time delay considered in this paper, each $\boldsymbol{\eta}_n$ needs to be estimated separately. Based on Cramér-Rao Bound (CRB) theorem from \cite{kay1993fundamentals}, the error covariance matrix of the unbiased estimator $\tilde{\boldsymbol{\eta}}_n$ of $\boldsymbol{\eta}_n$ satisfies
\begin{equation} \label{CRB_def}
    \mathbb{E}[(\tilde{\boldsymbol{\eta}}_n-\boldsymbol{\eta}_n)(\tilde{\boldsymbol{\eta}}_n-\boldsymbol{\eta}_n)^H]\geq \textbf{J}_{\boldsymbol{\eta}_n}^{-1}, 
\end{equation}
where $\textbf{J}_{\boldsymbol{\eta}_n}$ is the Fisher information matrix (FIM) and the CRB matrix is given by $CRB_{\boldsymbol{\eta}_n} = \textbf{J}_{\boldsymbol{\eta}_n}^{-1}$. For efficient expression, the signal model in \eqref{SensingsignalwithRIS} can be reformulated as
\begin{equation}\label{SensingsignalwithRIS2}
    y_{\text{S}}(t)= \boldsymbol{h}_{e}^{\text{tot}}(t,\boldsymbol{\eta})\boldsymbol{ \alpha} + n_k(t), \quad t\in[0,T_0],
\end{equation}
where $\boldsymbol{h}_{e}^{\text{tot}}(t,\boldsymbol{\eta}) =[h_e(t,\boldsymbol{\eta}_0), ...,h_e(t,\boldsymbol{\eta}_n), ..., h_e(t,\boldsymbol{\eta}_N)]$ with $h_e(t,\boldsymbol{\eta}_n) = e^{j2\pi f_{D_n}t}s_{k}(t-\tau_{n})$, and $\boldsymbol{\alpha} = [\alpha_0,...,\alpha_n,...,\alpha_N]^T$ as in \eqref{def_alpha_n}. 
The FIM $\textbf{J}_{\boldsymbol{\eta}_n} \in \mathbb{C}^{4\times 4}$ in \eqref{CRB_def} can be written as
\begin{equation} \label{FIM_total}
    \textbf{J}_{\boldsymbol{\eta}_n} = 
    \begin{bmatrix}
        J_{d_n,d_n} & J_{v_n,d_n}^T \\
        J_{v_n,d_n} & J_{v_n,v_n} 
    \end{bmatrix},
\end{equation}
where all elements are calculated in Appendix based on the signal model in \eqref{SensingsignalwithRIS2}. Finally, the CRB for range and velocity can be calculated as
\begin{align}
    CRB_{d_n} &= \left[\mathbf{J}_{\boldsymbol{\eta}_n}^{-1}\right]_{1,1}, \label{SensingRISCRB_d} \\
     CRB_{v_n} &= \left[\mathbf{J}_{\boldsymbol{\eta}_n}^{-1}\right]_{2,2}. \label{SensingRISCRB_v}
\end{align}

\section{RIS Deployment Problem Formulation} 
\label{form}
\begin{figure}[!t]
\centering
\includegraphics[width=2.8in,trim=20mm 10mm 0mm 0mm,clip]{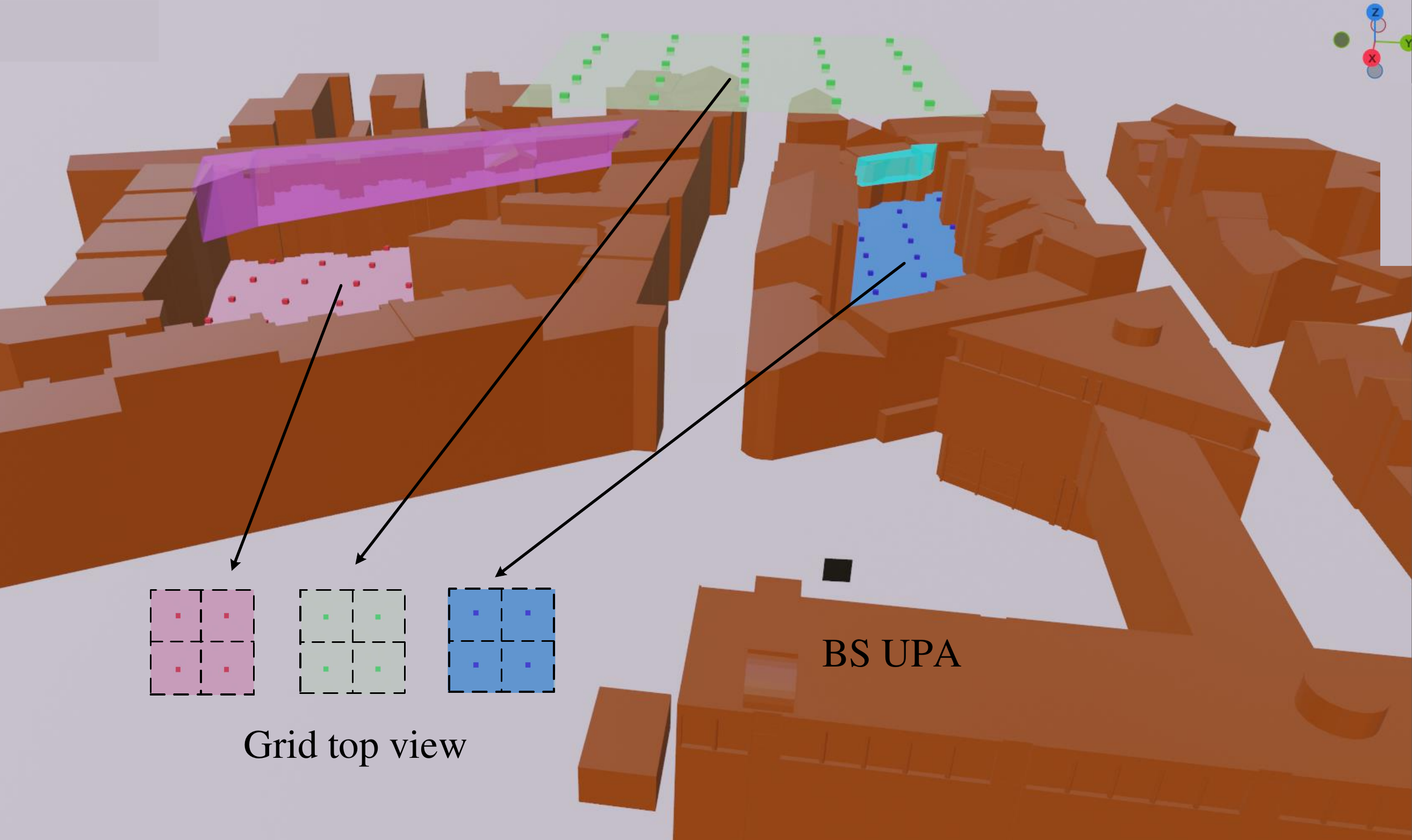}
\caption{The exemplary RIS-ISAC application scenario with $N=2$, based on the city environment around Kronenplatz in Karlsruhe, Germany. The colored notations are:
    \textcolor{black}{\rule{6pt}{6pt}}~BS UPA phase center position $\boldsymbol{p}_b$,
    \textcolor{red}{\rule{6pt}{6pt}}~UE grid center positions in set $\mathbf{S}_{1}$,
    \textcolor{red!60}{\rule{6pt}{6pt}}~RIS assisted area $\mathcal{R}_{1}^{\text{cov}}$,
    \textcolor{magenta!90}{\rule{6pt}{6pt}}~RIS deployable region $\mathcal{R}_{1}$;\;
    \textcolor{blue}{\rule{6pt}{6pt}}~UE grid center positions in set $\mathbf{S}_{2}$,
    \textcolor{blue!60!cyan}{\rule{6pt}{6pt}}~RIS-assisted area $\mathcal{R}_{2}^{\text{cov}}$,
    \textcolor{cyan}{\rule{6pt}{6pt}}~RIS deployable region $\mathcal{R}_{2}$;\;
    \textcolor{green}{\rule{6pt}{6pt}}~UAV grid center positions in set $\mathbf{S}_{u}$,
    \textcolor{green!40}{\rule{6pt}{6pt}}~UAV flight area $\mathcal{R}_{u}$.}
\label{RIS_ISAC_scenario}
\end{figure}

To meet the energy efficiency demands of the next-generation wireless communication networks, it is desirable to use RISs with the least possible unit cells, or equivalently, minium sizes, as the size of the RIS $A_n = A_uM_n$, where $A_u = D^2$ represents the unit cell size. Consequently, the RIS reflection gain $G_n(\theta_{n,k},\psi_{n,k})$ in \eqref{RIS2UE}, and the RIS incident gain $G_n(\theta_{n,b},\psi_{n,b})$ in \eqref{BS2RIS}, together yield an effective RIS gain $G_{\text{C},n} = G_n(\theta_{n,k},\psi_{n,k})G_n(\theta_{n,b},\psi_{n,b})$ for communication, which can be equivalently written as
\begin{equation} \label{RIS_gain}
    G_{\text{C},n} = \eta_{\text{C}}\cos\vartheta_{n,b}\cos\vartheta_{n,k} A_n^2 (\frac{4\pi}{\lambda_c^2})^2 \propto A_n^2,
\end{equation} 
following \cite{9837936}. The efficiency $\eta_{\text{C}}$ includes the effects of RIS reflection coefficients and unit cell radiation pattern for communication. The term $\vartheta_{n,b}$ is the incident angle from BS2RIS, and $\vartheta_{n,k}$ is the reflection angle from RIS2UE, as defined in Fig.~\ref{UPA_loc_coord}. Similarly, the effective RIS gain $G_{\text{S},n} = G_n(\theta_{n,u},\psi_{n,u})G_n(\theta_{n,b},\psi_{n,b})$ for sensing  can be equivalently represented as
\begin{equation} \label{RIS_gain_sensing}
    G_{\text{S},n} = \eta_{\text{S}}\cos\vartheta_{n,b}\cos\vartheta_{n,u} A_n^2 (\frac{4\pi}{\lambda_c^2})^2 \propto A_n^2.
\end{equation}
Here, the efficiency $\eta_{\text{S}}$ includes the effects of RIS reflection coefficients and unit cell radiation pattern for sensing, and $\vartheta_{n,u}$ is the reflection angle from RIS2UAV as defined in Fig.~\ref{UPA_loc_coord}. Obviously, \eqref{RIS_gain} and \eqref{RIS_gain_sensing} are embedded in \eqref{caschannel} and \eqref{SensingchannelwithRIS}, impacting the ISAC QoS. The communication SNR in \eqref{RIS_SINR} increases proportionally to the square of RIS size $A_n^2$, and the sensing CRBs in \eqref{SensingRISCRB_d}, \eqref{SensingRISCRB_v} are inversely proportional to $A_n^2$, as mentioned in Appendix. Therefore, to guarantee ISAC QoS requirements with minimum RIS size, optimization of the RIS deployment is necessary. This includes a comprehensive consideration incorporates environment,
spatial, and signal processing features to multiple RISs in high-fidelity scenarios at mmWave frequencies as introduced in the remainder of this section. To support clear description, an exemplary RIS-ISAC scenario is illustrated in Fig.~\ref{RIS_ISAC_scenario}.

\vspace{-0.5\baselineskip}
\subsection{RIS Position Constraints}\label{RISposition}
The deployment position of a RIS influences the attenuation factor, DoD and DoA of the cascaded channels in \eqref{caschannel} and \eqref{SensingchannelwithRIS}, which has to be carefully determined for RIS size reduction. First, to justify the areas that require RIS-assisted communication with moderate complexity, the entire ground layer is divided into rectangular UE grids utilizing the adaptive grid refinement (ARG) algorithm from \cite{long2025joint}. This algorithm ensures the fading factors of the channels to be relatively constant inside one grid, providing a representative UE SNR at the center position. By evaluating $\gamma_{\text{C},b,k}$ in \eqref{Only BS SINR} against the predefined threshold $\gamma^{\text{tr}}$, the set of UE grids that require RIS assistance can be derived, formulating the $n$-th RIS-assisted area $\mathcal{R}^{\text{cov}}_n$ in Fig.~\ref{RIS_ISAC_scenario}. Upon this, deploying RISs in the dense urban scenario also needs to strictly fulfill the constraints below:

\subsubsection{Building Attached RIS Deployment}
As noted in \cite{xu2020resource}, practical RISs are typically large panels, which may obstruct traffic or block sight, making them unsuitable for installation on streets or mounting on roadside infrastructure such as lamp posts. Therefore, the most promising deployment strategy for RIS is attaching to buildings, leveraging large free spaces and ensuring better coverage without side effects.

\subsubsection{Valid Link Access}
Besides building-attached deployment, to minimize RIS size and avoid redundant computation while ensuring ISAC QoS, the following link access conditions must be satisfied, which can be verified using IHE-RL mentioned in Section \ref{tool}. First, a LoS path between the RIS and BS must exist for proper illumination. Second, a LoS path between the RIS and the UAV in the air-layer area $\mathcal{R}_u$ in Fig.~\ref{RIS_ISAC_scenario} must exist for unbiased sensing. Third, there must be at least one path between the target UE grid center and a RIS with attenuation factor under $PL_{\text{max}}$ for valid signal reachability.

In summary, a problem to minimize the number of RIS panels while fulfilling the aforementioned conditions can be formulated and solved using existing methods, such as the greedy set cover algorithm \cite{young2008greedy}. As shown in Fig.~\ref{RIS_ISAC_scenario}, this yields the decision of $N$ RISs and the deployable region of every RIS $\mathcal{R}_n$. Each RIS assists $K_n$ UE grids in the set $\textbf{S}_n = \{\boldsymbol{p}_1,...,\boldsymbol{p}_{k_n},...,\boldsymbol{p}_{K_n}\}$, with $k_n \in \{1,2,...,K_n\}$, where $\boldsymbol{p}_{k_n}$ represents the center of the $k_n$-th UE grid. These grids form the $n$-th RIS-assisted area $\mathcal{R}^{\text{cov}}_n$, comprising the size $A_n^{\text{cov}}$. Similarly, the UAV flying area $\mathcal{R}_u$ is divided into $M_u$ UAV grids with center positions in the set $\textbf{S}_u = \{\boldsymbol{p}_{1}, \dots, \boldsymbol{p}_{m_u}, \dots, \boldsymbol{p}_{M_u}\}, m_u \in \{1,2,...,M_u\}$, with $\boldsymbol{p}_{m_u}$ representing the center position of the $m_u$-th UAV grid.

\vspace{-0.5\baselineskip}
\subsection{RIS 3D Orientation Model}\label{RISOri}
The orientation of RIS influences the signal reachability to the RIS-assisted area, the incident/reflection angles of the RIS in \eqref{RIS_gain} and \eqref{RIS_gain_sensing}. Therefore, properly aligned RIS orientation can effectively reduce the necessary RIS size while maintaining the desired ISAC QoS. Inspired by the movable antenna concept in \cite{zhu2023modeling}, the proposed RIS is mounted with a handle on the building side wall, with possible 2D orientation angles $(\theta_n^{\text{r}}, \psi_n^{\text{r}})$ as shown in Fig.~\ref{RIS_ori_setting}. Considering that a practical RIS is a one-side reflection panel, these orientations are defined within the ranges $\theta^{{\text{r}}}_n \in [\theta^{{\text{r}}}_{n,{\text{l}}}, \theta^{{\text{r}}}_{n,\text{h}}]$ and $\psi^{{\text{r}}}_n \in [\psi^{{\text{r}}}_{n,\text{l}},\psi^{{\text{r}}}_{n,\text{h}}]$ to ensure the reachability of RIS to target UE grids satisfying $PL_{\text{max}}$. Upon this, a general rotation matrix of RIS in the local coordinate can be expressed as
\cite{koehler1978euler}
\begin{equation}
\textbf{R}_{\text{tot}} = 
\begin{bmatrix}
\cos \psi^{\text{r}}_n & -\sin \psi^{\text{r}}_n & 0 \\
\sin \psi^{\text{r}}_n & \cos \psi^{\text{r}}_n & 0 \\
0 & 0 & 1
\end{bmatrix}
\begin{bmatrix}
\cos \theta_n^{\text{r}} & 0 & \sin \theta_n^{\text{r}} \\
0 & 1 & 0 \\
-\sin \theta_n^{\text{r}} & 0 & \cos \theta_n^{\text{r}}
\end{bmatrix}
.
\end{equation}
Let $\boldsymbol{e}_{\text{p}}$ denote the global coordinate unit vector of any RIS-related path, by calculating 
\begin{equation} \label{v_p_rot}
    [e_{n,p}^{\text{x}},e_{n,p}^{\text{y}},e_{n,p}^{\text{z}}]^T = \textbf{R}_{\text{tot}} \boldsymbol{e}_{\text{p}},
\end{equation}
and 
\begin{equation} \label{DoD/DoA_def}
    \theta_{n,p}= \arccos{e_{n,p}^{\text{z}}},\quad \psi_{n,p} = \arctan{(e_{n,p}^{\text{y}}/e_{n,p}^{\text{x}})},
\end{equation}
the RIS-related DoD/DoA $(\theta_{n,p},\psi_{n,p}), p\in\{b,k,u\}$ including RIS orientation could be derived, yielding the derivation of $(\theta_{n,k},\psi_{n,k})$ in \eqref{RIS2UE}, $(\theta_{n,b},\psi_{n,b})$ and $(\theta_{b,n},\psi_{b,n})$ in \eqref{BS2RIS}, and $(\theta_{n,u},\psi_{n,u})$ in \eqref{RIS2UAV}, respectively. Besides, \eqref{v_p_rot} is also utilized to calculate $\vartheta_{n,p} = \arccos{(e_{n,p}^{\text{x}})}$ in \eqref{RIS_gain} and \eqref{RIS_gain_sensing}.

\vspace{-0.5\baselineskip}
\subsection{RIS Beamforming Strategy} \label{RIS_bf_sec}
The RIS beamforming matrix $\boldsymbol{\Phi}_{n}$ dominates the ISAC QoS in \eqref{RxsignalwithRIS} and \eqref{SensingchannelwithRIS}, therefore, has to be considered as a deployment parameter. As stated in Section \ref{RISposition}, the $n$-th RIS supposes to cover $K_n$ UE grids in the set $\textbf{S}_n$. In practice, it is possible that one RIS might serve $K'_n$ UEs, satisfying $K'_n \leq K_n$ and $K'_n \leq K$, leading to an UE-number dependent performance. In this circumstance, the ideal RIS beamforming vector can be expressed by \cite{10438390}
\begin{equation} \label{RIS beamforming}
\boldsymbol{\epsilon}_n^{\text{ideal}}(\boldsymbol{\theta}_n',\boldsymbol{\psi}_n',\theta_{n,u},\psi_{n,u}) = \boldsymbol{\epsilon}_{\text{steer}}(\boldsymbol{\theta}_n',\boldsymbol{\psi}_n',\theta_{n,u},\psi_{n,u})+\boldsymbol{\epsilon}_{\text{dis}},
\end{equation}
where $\boldsymbol{\epsilon}_{\text{dis}}\in \mathbb{R}^{M_n \times 1}$ contains the distance phases introduced by different distances from the BS to each unit cell on the $n$-th RIS. 
The set $\boldsymbol{\theta}_n' = \{\theta_{n,1},...,\theta_{n,k'},...\theta_{n,K_n{'}}\}$ and $\boldsymbol{\psi}_n' = \{\psi_{n,1},...,\psi_{n,k'},...,\psi_{n,K'_n}\}$ with $k'\in\{ 1,2,..,K'_n\}$ in \eqref{RIS beamforming} contain the DoDs of the $n$-th RIS to communicate with the UEs. Together with the DoD to the UAV ($\theta_{n,u},\psi_{n,u}$), they determine the steering phase $\boldsymbol{\epsilon}_{\text{steer}} \in \mathbb{R}^{M_n\times 1}$ to form multi-beams for ISAC functionalities, with respect to the corresponding steering vectors $\boldsymbol{a}_y$ and $\boldsymbol{a}_z$ similar to \eqref{DoD_y} and \eqref{DoD_z} as
\begin{align} \label{multibeam_bf}
\boldsymbol{\epsilon}_{\text{steer}}(\boldsymbol{\theta}_n',\boldsymbol{\psi}_n',\theta_{n,u},\psi_{n,u}) = \notag\\ \angle[\sum_{k'=1}^{K'_n} \sqrt{\beta_{n,k'}} (\boldsymbol{a}_y(\theta_{n,k'},\psi_{n,k'}) &\otimes \boldsymbol{a}_z(\theta_{n,k'},\psi_{n,k'}))^* \notag\\ +
\sqrt{\beta_{n,u}} (\boldsymbol{a}_y(\theta_{n,u},\psi_{n,u}) &\otimes \boldsymbol{a}_z(\theta_{n,u},\psi_{n,u}))^{*}],
\end{align}
following \cite{li2014frequency} as a representative strategy. Here, $\beta_{n,k'}, \beta_{n,u} \in[0,1]$ are the RIS beamforming weighting factors for the $k'$-th UE and the UAV, satisfying $\sum_{k' = 1}^{K'_n}\beta_{n,k'} + \beta_{n,u} = 1$. After the phase quantization of $\boldsymbol{\epsilon}_n^{\text{ideal}}$ in \eqref{RIS beamforming} to $\epsilon_{n,m_n}$, the RIS beamforming matrix $\boldsymbol{\Phi}_{n}$ in \eqref{RxsignalwithRIS} is filled by the diagonal elements with $\boldsymbol{\phi}_n$. It is clearly visible in \eqref{multibeam_bf} that larger $K'_n$ leads to more computational effort. Nevertheless, to utilize reflectional RIS, the signals to different UEs need to be assigned with frequency division or time division to avoid interference \cite{li2021programmable}. Equivalently, these methods lead to the fact that multiple existing UEs mainly cause the received power reduction due to multi-beam signal scattering at the RIS, which could be compensated by adaptive BS transmit power control as widely used in industry \cite{benini2000survey}. Therefore, in this paper, it is assumed that one RIS serves $K'_n = 1$ UE to reduce the computation effort, and the BS transmits signal with the unit reference power, reducing \eqref{multibeam_bf} to a dual-beam scenario, making $\beta_{n,k'}$ unified to $\beta_{n,k}$, with the $k$-th UE lying at $\boldsymbol{p}_k \in \textbf{S}_n$. Considering $N$ RISs to optimize, weighting factors set for communication and sensing are denoted as $\boldsymbol{\beta}_{\text{C}}$ and $\boldsymbol{\beta}_{\text{S}}$, with each formed with every $\beta_{n,k}$ and $\beta_{n,u}$, respectively.

\vspace{-0.5\baselineskip}
\subsection{BS Beamforming Strategy} \label{BS_bf_sec}
BS beamforming determines the signal power allocated to each RIS, which directly influences the required RIS size to ensure sufficient ISAC QoS. An effective BS-related channel matrix $\textbf{H}_{b} \in \mathbb{C}^{(K+1)\times M_b}$ between the UAV and all $K$ UEs is
\begin{equation}\label{BSfullmatrix}
    \textbf{H}_{b}=[\textbf{h}_{b,u}^T,\textbf{H}_{b}^{\text{RIS}},\textbf{H}_{b}^{\text{UE}}]^T,
\end{equation}
with the sensing incident channel $\textbf{h}_{b,u}$ in \eqref{BStoUAVsensing}. The matrix $\textbf{H}_{b}^{\text{RIS}}\in\mathbb{C}^{M_b\times N}$ contains the RIS-assisted channel to $N$ UEs gathering $\boldsymbol{h}_{\text{C},b,n,k}$ in \eqref{caschannel}. The matrix $\textbf{H}_{b}^{\text{UE}} \in\mathbb{C}^{M_b\times K_b}$ contains $K_b$ BS2UE channels $\boldsymbol{h}_{\text{C},b,k}$ from \eqref{BSmultipathchannelsingleUE} that could be directly served by BS with sufficient SNR, having $N + K_b = K$. 
Considering BS beamforming strategy in a path-based model logic and proper interference cancellation, singular value decomposition (SVD) is a preferred representative method. As proposed in Section \ref{RISposition}, since a RIS always has LoS access to the BS, each MIMO matrix between BS and RIS has a dominant singular value. Therefore, the SVD-based beamforming matrix at the BS could be expressed as
\begin{equation}\label{BSISACBF}
    \textbf{W}_{\text{SVD}}=[\boldsymbol{w}_{b,u},\textbf{W}_{b}^{\text{RIS}},\textbf{W}_{b}^{\text{UE}}],
\end{equation}
with $\boldsymbol{w}_{b,u}$ from \eqref{def_alpha_n}, $\textbf{W}_{C}$ from \eqref{RxsignalonlyBS} reformed as $\textbf{W}_{C} = [\textbf{W}_{b}^{\text{RIS}},\textbf{W}_{b}^{\text{UE}}]$, grouping the BS2RIS beamforming matrix as $\textbf{W}_{b}^{\text{RIS}}$ and the BS2UE beamforming matrix as $\textbf{W}_{b}^{\text{UE}}$. Considering power allocation to each beamforming vector, the final BS beamforming matrix becomes
\begin{equation}\label{ISACpower}
    \textbf{W}_{\text{ISAC}} = \textbf{W}_{\text{SVD}} \mathbf{\Omega}_{\text{ISAC}}, 
\end{equation}
with $\mathbf{\Omega}_{\text{ISAC}}=\text{diag}(\sqrt{\omega_0},...,\sqrt{\omega_k},...,\sqrt{\omega_K})$ satisfying $\text{Tr}\{\mathbf{\Omega}_{\text{ISAC}}^2\}=1$. 

\vspace{-0.5\baselineskip}
\subsection{Size-to-Coverage Sum Ratio Minimization} \label{problem_min}
Summarizing the aforementioned properties and constraints, an optimization problem of RIS deployment is formulated as 
\begin{align}
\tag{P0}\label{prob:P0}
&\min_{\textbf{P},\boldsymbol{\Theta}^{\text{r}},\boldsymbol{\Psi}^{\text{r}},\boldsymbol{\beta}_{\text{C}},\boldsymbol{\beta}_{\text{S}},\mathbf{\Omega}_{\text{ISAC}}} \sum_{n = 1}^{N} \frac{A_n}{A_n^{\text{cov}}} \\
\text{s.t.} \notag \\
&\text{C1:} \quad \gamma_{\text{C},b,n,k} \geq \gamma^{\text{tr}}, \quad \forall \boldsymbol{p}_{k}\in \textbf{S}_n,\notag \\
&\text{C2:} \quad CRB_{d_n} \leq CRB_{d^{\text{tr}}}, CRB_{v_n} \leq CRB_{v^{\text{tr}}}, \boldsymbol{p}_u\in \textbf{S}_u, \notag \\
&\text{C3:} \quad \boldsymbol{p}_n \in \mathcal{R}_n, \notag \\
&\text{C4:} \quad \theta^{\text{r}}_n \in [\theta^{\text{r}}_{n,{\text{l}}},\theta^{\text{r}}_{n,\text{h}}], \psi^{\text{r}}_n \in [\psi^{\text{r}}_{n,\text{l}},\psi^{\text{r}}_{n,\text{h}}], \notag \\
&\text{C5:} \quad \epsilon_{n,m_n} \in \{\frac{2\pi l}{2^L}|l = 0,1,...,2^L-1\}, \forall m_n\in[1,M_n],\notag\\
&\text{C6:} \quad \text{Tr}\{\mathbf{\Omega}_{\text{ISAC}}^2\}=1,\notag\\
&\text{C7:} \quad \beta_{n,k} + \beta_{n,u}=1. \notag
\end{align}
 
For a better relation to real-world environments and energy-efficiency-driven expression, the objective function is to minimize the additive ratio between the size of every RIS $A_n$ and its assisting area $A_n^{\text{cov}}$, namely multi-RIS size-to-coverage sum ratio in this paper. The optimization is realized by adjusting the parameters including RIS positions $\textbf{P} = \{\boldsymbol{p}_1,...,\boldsymbol{p}_n,...,\boldsymbol{p}_N\}$, orientations $\mathbf{\Theta}^{\text{r}} = [\theta_{1}^\text{r},...,\theta_n^\text{r},...,\theta_N^\text{r}]$, $\mathbf{\Psi}^{\text{r}} = [\psi_{1}^\text{r},...,\psi_n^\text{r},...,\psi_N^\text{r}]$, RIS beamforming strategy and BS ISAC power allocation associated with $\boldsymbol{\beta}_{\text{C}}$, $\boldsymbol{\beta}_{\text{S}}$ and $\mathbf{\Omega}_{\text{ISAC}}$. For constraints, C1 guarantees all the UE grids covered with sufficient communication SNR. For BS served UE, it is evaluated with \eqref{Only BS SINR}. For RIS-assisted UE, it is evaluated with \eqref{RIS_SINR}, with interference not further considered since the interference path would be either blocked by obstacles, or suffer from stronger cascaded loss from undesired RIS with relative weak beamforming; C2 ensure that each BS2UAV path and RIS2UAV path satisfies the CRB requirements in Section \ref{Sensing Model}; C3 defines the deployable region of RISs in Section \ref{RISposition}; C4 denotes the RIS orientation range to meet UE reachability demand in Section \ref{RISOri}; C5 defines the $L$-bit discrete RIS phase shift resolution of \eqref{RIS beamforming}; C6 constraints the normalized power at BS in \eqref{ISACpower}; C7 limits the beamforming weighting factors of the RISs in \eqref{multibeam_bf}.

\section{Proposed Algorithm} \label{proposed_algo}
The problem \ref{prob:P0} is a non-convex mixed-integer nonlinear program mapping to real-world environments, which is complicated to solve with multiple variables and constraints. Therefore, in this section, \ref{prob:P0} is firstly simplified, and then solved by a two-step iterative algorithm.

\subsection{Problem Simplification} \label{problem_simp}
Deriving the RIS beamforming matrix $\boldsymbol{\Phi}_{n}$ with more than ten thousand unit cells \cite{xu2020resource} for the grid-based simulation model, which addresses practical environments, makes \ref{prob:P0} extremely time-consuming to solve. Therefore, an equivalent gain scaling method is applied, instead of explicitly computing the entire RIS beamforming matrix for ISAC QoS evaluation. As discussed in \eqref{RIS_gain}, the RIS-assisted communication SNR is proportional to $A^2_n = (A_u{M_n})^2$. Under normalized BS power and RIS beamforming weighting factor, the reference SNR $\gamma_{\text{C},b,n,k}^{\text{ref}}$ for the $k$-th UE at any grid position $\boldsymbol{p}_{k}\in \textbf{S}_n$ can be computed using a RIS with a reasonable reference number of unit cells ${M^{\text{ref}}}$ following \eqref{RIS_SINR}, and scaled with the square of the factor $\alpha_n = \frac{{M_n}}{{M^{\text{ref}}}}$ to account for changes in RIS size. Furthermore, since the required transmit power for the BS2UE path can be determined by evaluating $\gamma_{\text{C},b,k}$ in \eqref{Only BS SINR} against $\gamma^{\text{tr}}$, which is independent of RIS assistance, it can be excluded from the optimization. This results in the transformation of $\mathbf{\Omega}_{\text{ISAC}}$ in \eqref{ISACpower} to $\mathbf{\Omega}_{\text{ISAC}}'=\text{diag}(\sqrt{\omega_0},...,\sqrt{\omega_n},...,\sqrt{\omega_N})$. Then, by taking the RIS beamforming weighting factors from C7 into consideration, the SNR in \eqref{RIS_SINR} is equivalent to
\begin{equation} \label{UE_snr_with_ris_size_w}
    \gamma_{\text{C},b,n,k} = \beta_{n,k}\omega_n\alpha_n^2 \gamma_{\text{C},b,n,k}^{\text{ref}}.
\end{equation}
The same logic goes for UAV sensing, yielding the CRBs from \eqref{SensingRISCRB_d} and \eqref{SensingRISCRB_v} based on the reference boundaries $CRB_{d_n}^{\text{ref}}$ and $CRB_{v_n}^{\text{ref}}$ as
\begin{align} 
    CRB_{d_n} = \frac{CRB_{d_n}^{\text{ref}}}{(1-\beta_{n,k})\omega_n\alpha_n^2}, \label{CRB_with_ris_size_w_1} \\
    CRB_{v_n} = \frac{CRB_{v_n}^{\text{ref}}}{(1-\beta_{n,k})\omega_n\alpha_n^2}. \label{CRB_with_ris_size_w_2}
\end{align}
By substituting \eqref{UE_snr_with_ris_size_w}, \eqref{CRB_with_ris_size_w_1} and \eqref{CRB_with_ris_size_w_2} into constraints C1, C2 considering $n\neq 0$ in \ref{prob:P0}, each of them leads to $\omega_n \alpha_n^2 \geq c_{n,i}, i=\{1,2,3\}$ with
\begin{align} 
    c_{n,1} &= \frac{\gamma^{\text{tr}}}{\beta_{n} {\gamma_{\text{C},b,n}^{\text{ref}}}}, \label{constraint_simp_1}\\
    c_{n,2} &= \frac{CRB_{d_n}^{\text{ref}}}{(1-\beta_{n})CRB_{d^{\text{tr}}}}, \label{constraint_simp_2}\\
    c_{n,3} &= \frac{CRB_{v_n}^{\text{ref}}}{(1-\beta_{n})CRB_{v^{\text{tr}}}}.
    \label{constraint_simp_3}
\end{align}
By defining $\mathbf{\Gamma}_{\text{C},b,n,K_n}^{\text{ref}}$ as the set of $\gamma_{\text{C},b,n,k}^{\text{ref}}$ derived at $K_n$ possible UE grid center positions of $\boldsymbol{p}_{k} \in \textbf{S}_n$ as in Fig.~\ref{RIS_ISAC_scenario}, $\gamma_{\text{C},b,n}^{\text{ref}} = \min(\mathbf{\Gamma}_{\text{C},b,n,K_n}^{\text{ref}})$ in \eqref{constraint_simp_1} is the worst SNR conditioned grid for the $n$-th RIS. Let $k^{\star}$ be the UE grid index where this minimum is attained, $\beta_{n,k^{\star}}$ for this specific grid is denoted as $\beta_n$ in \eqref{constraint_simp_1}-\eqref{constraint_simp_3} for simplicity. Up to now, the problem is transformed to satisfy the SNR requirement of the worst-case RIS-assisted UE grid. Then, to jointly meet C1 and C2, the most strict constraint is selected by defining $c_n = \max(c_{n,i})$, leading to the lower bound of RIS size fulfilling $\alpha_n \geq \sqrt{c_n/\omega_n}$, expressed as
\begin{equation} \label{RIS_size_cal}
    A_n=\frac{\sqrt{c_n} A_u M^{\text{ref}}}{\sqrt{\omega_n}}.
\end{equation}

Afterwards, the constraint C2 when $n=0$ is the CRB requirements for the BS2UAV path independent to RIS, with $\omega_0$ can be calculated based on the UAV position, further transforming C6 to $\sum_{n=1}^{N} \omega_n = 1 - \omega_0$. Finally, concluding all the aforementioned aspects, problem \ref{prob:P0} is equivalent to 
\begin{equation} 
\tag{P1}\label{prob:P1}
\min_{\textbf{P},\boldsymbol{\Theta}^{r},\boldsymbol{\Psi}^{r},\boldsymbol{\beta}_\text{C},\mathbf{\Omega}_{\text{ISAC}}'} \sum_{n=1}^{N} \frac{\sqrt{c_n} A_u M^{\text{ref}}}{\sqrt{\omega_n} \cdot A_n^{\text{cov}}} 
\end{equation}
s.t.:
\begin{align}
    &\sum_{n=1}^{N} \omega_n = 1 - \omega_0, \label{p_sum_constraint}\\
    &\boldsymbol{p}_u\in \textbf{S}_u, \label{uav_constraint}\\
    &\text{and C3, C4, C5}. \notag
\end{align}
As the UAV position is confined to $\textbf{S}_u$, the leftover of C2 remains as \eqref{uav_constraint}. 

\vspace{-0.5\baselineskip}
\subsection{Two-Step Iterative Algorithm} 
\label{fixed_pos_algo}
To solve \ref{prob:P1}, the proposed algorithm processes two steps iteratively. Starting with an initial random RIS position set, \textit{Step 1} determines the optimization parameters and objective values at the specific RIS positions. Based on the derived objective values, \textit{Step 2} provides new position candidates to update the position set for \textit{Step 1}, approaching the position yielding the minimum multi-RIS size-to-coverage sum ratio. The details are introduced as follows:

\subsubsection{Step 1: RIS Deployment Determination at Specific Positions}
\label{fixed_pos_algo1}
Based on the given RIS deployment position set (initially random and provided by \textit{Step 2} for later iterations), \textit{Step 1} begins with the determination of the RIS 3D orientations. As mentioned in \eqref{RIS_gain} and \eqref{RIS_gain_sensing}, the communication SNR in \eqref{RIS_SINR} is proportional to $\cos\vartheta_{n,b}\cos\vartheta_{n,k}A_n^2$, and the sensing CRBs in \eqref{SensingRISCRB_d} and \eqref{SensingRISCRB_v} are inversely proportional to $\cos\vartheta_{n,b}\cos\vartheta_{n,u}A_n^2$. To evaluate the effective reference size of the $n$-th RIS, a factor is defined as
\begin{align} \label{A_tot}
A_n^{\text{avg}} \!= \!{(A_uM^{\text{ref}})}^2\!\cos\!\vartheta_{n,b} \!
\left(\! \frac{1}{K_n} \!\sum_{k=1}^{K_n} \!\cos\!\vartheta_{n,k}\! +\! \frac{1}{M_u} \!\sum_{u=1}^{M_u} \!\cos\!\vartheta_{n,u}\! \right).
\end{align}
The idea is to average the orientation contributions over all the grids for UE communication and UAV sensing considering fairness. The two contributions are equally summed up in the brackets defining the same importance of communication and sensing, which could be also adjusted based on specific applications. By maximizing each $A_n^{\text{avg}}$, the orientations $\tilde{\mathbf{\Theta}}^{\text{r}}(\textbf{P}) = [\tilde{\theta}_{1}^{\text{r}},...,\tilde{\theta}_{n}^{\text{r}},...,\tilde{\theta}_{N}^{\text{r}}]$ and $\tilde{\mathbf{\Psi}}^{\text{r}}(\textbf{P}) =[\tilde{\psi}_{1}^{\text{r}},...,\tilde{\psi}_{n}^{\text{r}},...,\tilde{\psi}_{N}^{\text{r}}]$ can be determined under C5, yielding the largest effective reference size for RISs at their specific positions $\textbf{P}$.

Next, for each UAV position, the minimum RIS size will be determined. The obstacle is that, the term $c_{n,i}$ in \eqref{constraint_simp_1}-\eqref{constraint_simp_3} is still related to the RIS power weighting factor $\beta_{n}$, yielding non-convex and non-smooth behaviors of \ref{prob:P1}. As a solution, a set of $\beta_{n}\in\mathbf{B}_{\text{C}}$ could be defined, e.g., using hierarchical set or any other set-construction methods, to support the computation of the corresponding $c_{n,i}$. The reason that such computation could be desirable is because that another key optimization parameter $\omega_n$ in \ref{prob:P1} can be expressed in closed-form utilizing the Karush-Kuhn-Tucker (KKT) conditions \cite{ruszczynski2011nonlinear} as follows. The Lagrangian function of \ref{prob:P1} is defined by introducing a dual variable $\lambda$ for the equality constraint in \eqref{p_sum_constraint}
\begin{align}
    \mathcal{L}&(\omega_0,\omega_1, \ldots,\omega_n,\ldots \omega_{N}, \lambda) = \notag\\
    &\sum_{n=1}^{N} \frac{\sqrt{c_n} A_u M^{\text{ref}}}{\sqrt{\omega_n}\cdot A_n^{\text{cov}}}
    + \lambda \left( \sum_{n=1}^{N} \omega_n + \omega_0 -1 \right).
\end{align}
According to the KKT conditions, the stationary condition requires that the partial derivative of the Lagrangian with respect to each $\omega_n$ vanishes at the optimal point with
\begin{equation}
    \frac{\partial \mathcal{L}}{\partial \omega_n} = -\frac{\sqrt{c_n} A_u M^{\text{ref}}}{2\omega_n^{3/2}A_n^{\text{cov}}} + \lambda = 0,
\end{equation}
which leads to $\omega_n = {(\frac{\sqrt{c_n}A_u M^{\text{ref}}}{2A_n^{\text{cov}} \lambda})}^{2/3}$. By Substituting $\omega_n$ into the constraint \eqref{p_sum_constraint}, it yields
\begin{equation}
    \sum_{n=1}^N {(\frac{\sqrt{c_n}A_u M^{\text{ref}}}{2A_n^{\text{cov}} \lambda})}^{2/3} = 1 - \omega_0.
\end{equation}
Finally, solving $\lambda$ and substituting it back to the expression for $\omega_n$ gives the optimal solution
\begin{equation} \label{rho_final}
\omega_n =   
\frac{{(\frac{\sqrt{c_n}A_u M^{\text{ref}}}{2A_n^{\text{cov}}})}^{2/3}}{\sum_{n'=1}^N {(\frac{\sqrt{c_{n'}}A_u M^{\text{ref}}}{2A_{n'}^{\text{cov}}})}^{2/3}}\cdot (1 - \omega_0).
\end{equation}
Considering $\boldsymbol{\beta}_{\text{C}} = [\beta_1,...,\beta_n,...,\beta_N]$ after the simplification in Subsection \ref{problem_simp} with each $\beta_n$ traversing $\mathbf{B}_{\text{C}}$, a set $\mathbf{B}_{\text{C}}^{\text{tot}}$ is defined as all possible collections of $\boldsymbol{\beta}_{\text{C}}$. Until now, all the optimization variables in \ref{prob:P1} are incrementally determined. Therefore, $E(\textbf{P}) = \sum_{n=1}^{N} \frac{\sqrt{c_n} A_u M^{\text{ref}}}{\sqrt{\omega_n} \cdot A_n^{\text{cov}}}$ can be calculated at RIS positions $\textbf{P}$ using different $\boldsymbol{\beta}_{\text{C}}$ collections, and the $\tilde{\boldsymbol{\beta}}_{\text{C}}(\textbf{P})= [\tilde{\beta}_1,...,\tilde{\beta}_n,...,\tilde{\beta}_N]$ that results in the minimum $E(\textbf{P})$ can be simply detected. 
Therefore, each corresponding $\tilde{\omega}_n$ and $A_n$ can be calculated using \eqref{rho_final} and \eqref{RIS_size_cal}, forming $\tilde{\boldsymbol{\Omega}}'_{\text{ISAC}}(\textbf{P}) = \text{diag}(\sqrt{\tilde{\omega}_1},...,\sqrt{\tilde{\omega}_n},...,\sqrt{\tilde{\omega}_N})$ and RIS size set $\textbf{A}(\textbf{P}) = [A_1,...,A_n,...,A_N]$, respectively. By repeating this procedure for all the UAV grids in \eqref{uav_constraint}, and picking the maximum RIS size $\tilde{A}_{n}$, the CRBs for the UAV at any interested position are ensured. Then, an objective value $\tilde{E}(\textbf{P}) =\sum_{n=1}^{N}\tilde{A}_n/A_n^{\text{cov}}$ can be derived, thereby forming the output RIS sizes set $\tilde{\textbf{A}}(\textbf{P}) = [\tilde{A}_1,...,\tilde{A}_n,...,\tilde{A}_N]$.

\subsubsection{Step 2: RIS Position Set Determination}\label{NM}

The aforementioned derivation of $\tilde{E}(\textbf{P})$ provides a local optimum based on the given RIS position set, yielding the necessity for \textit{Step 2} to match global optimum position. To find an appropriate method, it is important to note that evaluation based on a realistic scenario does not follow typical statistical behavior, making the channel model more complex and irregular, thus limiting the applicability of gradient-based methods commonly used for statistical studies. Additionally, the QoS evaluation of the multiplex network, where ISAC and RIS introduce multi-matrices and significantly increase the number of system nodes, requires high computational costs. After investigation, the Nelder-Mead (NM) algorithm is identified as an effective, reliable, and fast-converging optimization method with geometrical sampling property, making it well-suited for solving the proposed RIS deployment problem \cite{liu2024enhanced}. The detailed application of the NM method in this paper is presented below.

In general, the idea is to examine the RIS deployable region $\mathcal{R}_n$ and update RIS position set for \textit{Step 1}, approaching the global optimum after each iteration. Initially, when iteration index $q = 0$, $M_s+1$ positions for each RIS are randomly chosen within $\mathcal{R}_n$, constructing the position set $\mathbf{S}_q = \{\textbf{P}_{(1,q)},...,\textbf{P}_{(m_s,q)},...,\textbf{P}_{(M_s,q)},\textbf{P}_{(M_s+1,q)}\}, m_s\in\{1,2,...,M_s+1\}$, with $\textbf{P}_{(m_s,q)}=\{\boldsymbol{p}_{1,(m_s,q)},...,\boldsymbol{p}_{n,(m_s,q)},...,\boldsymbol{p}_{N,(m_s,q)}\}$ representing the set of $N$ RIS positions. 
At the $q$-th iteration, $\textbf{P}_{(m_s,q)}$ in $\mathbf{S}_q$ is first sorted in ascending order according to the objective value $\tilde{E}(\textbf{P}_{(m_s,q)})$ derived in \textit{Step 1} in a position and iterative indexed format, yielding an ordered set $\mathbf{S}_q'\! =\! \{\textbf{P}_{(1,q)}',...,\!\textbf{P}_{(m_s,q)}',...,\!\textbf{P}_{(M_s,q)}', \!\textbf{P}_{(M_s+1,q)}'\}$. 
Afterwards, the worst positions $\textbf{P}_{(M_s+1,q)}'$ will be updated by 
\begin{align} \label{NM_update}
    \textbf{P}_{(M_s+1,q+1)} = (1+\mu){\bar{\textbf{P}}}_{(q)}'-\mu\textbf{P}_{(M_s+1,q)}',
\end{align}
where $\bar{\textbf{{P}}}_{(q)}' = \frac{1}{M_s}\sum_{m_s = 1}^{M_s}\textbf{P}_{(m_s,q)}'$ denotes the centroid of the $M_s$ positions excluding the worst position. Based on the current search pattern, $\mu$ varies correspondingly, details of which can be found in \cite{long2025joint}. Then, $\textbf{S}_{q+1} = \{\textbf{P}_{(1,q)}',..., \textbf{P}_{(m_s,q)}',...,\textbf{P}_{(M_s,q)}',\textbf{P}_{(M_s+1,q+1)}\}$ is formed for the next iteration as an updated set for \textit{Step 1}. 
Defining
\begin{equation} \label{conv_dist_set}
\textbf{D}_{n,q} \!=\! \left\{\! \left\| \boldsymbol{p}_{n,(i,q)} - \boldsymbol{p}_{n,(j,q)} \right\| \;\middle| i,j \!\in\! \{1,2, \ldots, M_s+1\}, i \neq j \right\},
\end{equation}
which is the pairwise distance set between the positions of the $n$-th RIS across all $M_s+1$ positions in the $q$-th iteration. When $\max(\textbf{D}_{n,q}) \leq d_{\min},\forall n$ holds, all the $M_s+1$ positions for every $n$-th RIS lie within a small sphere with diameter $d_{\min}$, making further improvement less probable. Then, the NM has converged and the optimized RIS positions $\textbf{P}_{(1,q)}'$ in $\textbf{S}_q'$ result in the minimum $\tilde{E}(\textbf{P}_{(1,q)}')$ is derived.

\vspace{-0.5\baselineskip}
\subsection{Algorithm Overview}

The entire flow of the proposed algorithm to solve \ref{prob:P1} for the minimization of multi-RIS size-to-coverage sum ratio is presented as Alg.~\ref{algo_1}. The input of the algorithm are the scenario-related parameters and constrains introduced in the previous sections. By iteratively operating \textit{Step 1} and \textit{Step 2}, optimized RIS positions $\textbf{P}_{\text{opt}}$, orientations $\mathbf{\Theta}^{\text{r}}_{\text{opt}}$, $\mathbf{\Psi}^{\text{r}}_{\text{opt}}$, and sizes ${\textbf{A}_{\text{opt}}}$ are derived. Besides, the RIS beamforming weighting factor set $\tilde{\boldsymbol{\beta}}_{\text{C,opt}}$ and ISAC beamforming set $\tilde{\boldsymbol{\Omega}}'_{\text{ISAC,opt}}$ for all the UAV grids are returned. Particularly, the $\textbf{P}_{(m_s,q)}$ in Alg.~\ref{algo_1} refers to the specific position $\textbf{P}$ from \textit{Step 1} indexed by position and iteration. For initialization, RIS deployment position set $\textbf{S}_0$ is randomly chosen in line 1. First, the orientations $\tilde{\mathbf{\Theta}}^{\text{r}}(\textbf{P}_{(m_s,q)})$ and $\tilde{\mathbf{\Psi}}^{\text{r}}(\textbf{P}_{(m_s,q)})$ are determined to maximize \eqref{A_tot} in line 3. Then, regarding UAV at $\boldsymbol{p}_{m_u}$, line 4 to 16 are processed. For each $\boldsymbol{\beta}_{\text{C}} \in\mathbf{B}_{\text{C}}^{\text{tot}}$, $c_n = \max(c_{n,i})$ is calculated using \eqref{constraint_simp_1}-\eqref{constraint_simp_3} in line 7. Afterwards, $\omega_n$ is determined using \eqref{rho_final} in line 8 and $E(\textbf{P}_{(m_s,q)})$ is derived in line 10. Then, for each $m_u$-th UAV, $\tilde{\boldsymbol{\beta}}_{\text{C}}(\textbf{P}_{(m_s,q)})$ that minimizes $E(\textbf{P}_{(m_s,q)})$ obtained in \textit{Step 1} is saved in a storage matrix $\tilde{\boldsymbol{\beta}}_{\text{C},M_u}(\textbf{P}_{(m_s,q)})$, and its corresponding $\tilde{\boldsymbol{\Omega}}'_{\text{ISAC}}(\textbf{P}_{(m_s,q)})$ is derived and saved in a storage matrix $\tilde{\boldsymbol{\Omega}}'_{\text{ISAC},M_u}(\textbf{P}_{(m_s,q)})$ from line 12-14. Then, the RIS size set $\textbf{A}(\textbf{P}_{(m_s,q)})$ can be obtained using \eqref{RIS_size_cal} in line 15. After the aforementioned loop, the final RIS size set $\tilde{\textbf{A}}(\textbf{P}_{(m_s,q)})$ is selected with its element $\tilde{A}_n$ as the maximum across all the $M_u$ UAV grids in line 17. Accordingly, $\tilde{E}(\textbf{P}_{(m_s,q)})$ is obtained, and $\textbf{S}_q$ can be ordered in line 19. Next, the worst positions $\textbf{P}_{(M_s+1,q+1)}$ calculated using \eqref{NM_update} in line 21 will be replaced and $\textbf{S}_{q+1}$ can be formed in line 22 for the next iteration. Once the convergence condition is fulfilled in line 20, the output could be determined for the RIS deployment optimization.

\begin{algorithm}
 \caption{RIS size minimization algorithm}
 \begin{algorithmic}[1]
 \renewcommand{\algorithmicrequire}{\textbf{Input:}}
 \renewcommand{\algorithmicensure}{\textbf{Output:}}
 \REQUIRE $\boldsymbol{p}_b$, $f_c$, $A_n^{\text{cov}}$, $N$, $M_u$, $K_n$, $L$, $\textbf{S}_n$, $\mathcal{R}_n$, $\textbf{S}_u$, $A_u$, ${M^{\text{ref}}}$, $\gamma^{\text{tr}}$, $CRB_{d^{\text{tr}}}$, $CRB_{v^{\text{tr}}}$, $\theta^{\text{r}}_{n,{\text{l}}}$, $\theta^{\text{r}}_{n,\text{h}}$, $\psi^{\text{r}}_{n,\text{l}}$, $\psi^{\text{r}}_{n,\text{h}}$, $M_s$, $\mathbf{B}_{\text{C}}^{\text{tot}}$
    \STATE \textbf{Initialization}: Form random RIS position set $\textbf{S}_0$ 
        \FORALL {$m_s\in [1,M_s+1]$ at $\textbf{P}_{(m_s,q)}$}
            \STATE  Obtain $\tilde{\mathbf{\Theta}}^\text{r}(\textbf{P}_{(m_s,q)}), \tilde{\mathbf{\Psi}}^\text{r}(\textbf{P}_{(m_s,q)})$ by maximizing \eqref{A_tot} 
            \FORALL {$m_u \in [1,M_u]$ with UAV at $\boldsymbol{p}_{m_u}$}
                 \FORALL {$\boldsymbol{\beta}_{\text{C}}\in \mathbf{B}_{\text{C}}^{\text{tot}}$}
                     \FORALL {$n\in [1,N]$ with $n$-th RIS at $\boldsymbol{p}_{n,(m_s,q)}$} 
                        \STATE Derive $c_n = \max(c_{n,i})$ using \eqref{constraint_simp_1}-\eqref{constraint_simp_3}
                        \STATE Derive $\omega_n$ using \eqref{rho_final}
                    \ENDFOR
                    \STATE Derive $E(\textbf{P}_{(m_s,q)})$
                \ENDFOR
                \STATE Obtain $\tilde{\boldsymbol{\beta}}_{\text{C}}(\textbf{P}_{(m_s,q)})$ by minimizing $E(\textbf{P}_{(m_s,q)})$
                \STATE Save $\tilde{\boldsymbol{\beta}}_{\text{C}}(\textbf{P}_{(m_s,q)})$ to $\tilde{\boldsymbol{\beta}}_{\text{C},M_u}(\textbf{P}_{(m_s,q)})$ 
                \STATE Derive $\tilde{\boldsymbol{\Omega}}_{\text{ISAC}}'(\textbf{P}_{(m_s,q)})$ and save to $\tilde{\boldsymbol{\Omega}}'_{\text{ISAC},M_u}(\textbf{P}_{(m_s,q)})$
                \STATE Derive corresponding $\textbf{A}(\textbf{P}_{(m_s,q)})$ using \eqref{RIS_size_cal}
            \ENDFOR
            \STATE Obtain maximum $\tilde{\textbf{A}}(\textbf{P}_{(m_s,q)})$ among $M_u$ UAV grids
        \ENDFOR
         \STATE Derive $\tilde{E}(\textbf{P}_{(m_s,q)})$ and order $\textbf{S}_q$ into $\textbf{S}_q'$ 
         \IF{$\exists n,\max (\textbf{D}_{n,q}) > d_{\min}$}
            \STATE Derive $\textbf{P}_{(M_s+1,q+1)}$ using \eqref{NM_update} 
            \STATE $\textbf{S}_{q+1} = \{\textbf{P}_{(1,q)}',..., \textbf{P}_{(m_s,q)}',...,\textbf{P}_{(M_s,q)}',\textbf{P}_{(M_s+1,q+1)}\}$
            \STATE $q = q+1$ and go to line 2
        \ELSE
            \STATE Break and output
        \ENDIF
    \ENSURE $\textbf{P}_{\text{opt}} = \textbf{P}_{(1,q)}'$, ${\textbf{A}_{\text{opt}}} =\tilde{\textbf{A}}(\textbf{P}'_{(1,q)})$, $\mathbf{\Theta}^{\text{r}}_{\text{opt}}=\tilde{\mathbf{\Theta}}^\text{r}(\textbf{P}'_{(1,q)})$, $\mathbf{\Psi}^\text{r}_{\text{opt}}=\tilde{\mathbf{\Psi}}^\text{r}(\textbf{P}'_{(1,q)})$, $\tilde{\boldsymbol{\beta}}_{\text{C,opt}} = \tilde{\boldsymbol{\beta}}_{\text{C},M_u}(\textbf{P}'_{(1,q)})$ and $\tilde{\boldsymbol{\Omega}}'_{\text{ISAC,opt}} = \tilde{\boldsymbol{\Omega}}'_{\text{ISAC},M_u}(\textbf{P}'_{(1,q)})$
 \end{algorithmic} 
 \label{algo_1}
\end{algorithm}

\section{Simulation results and analysis} \label{sim}
In this section, the simulation settings will be introduced. Then, the performance analysis of the RIS deployment, and the comparison to previous studies are presented.

\vspace{-0.5\baselineskip}
\subsection{Simulation settings}
The simulation settings are listed in Table~\ref{sim_set}, with the key ones highlighted as follows. The reference size, radiation pattern, and reflection coefficient of the RIS are adopted from the $M^{\text{ref}} = 20 \times 20$ RIS in \cite{wan2021reconfigurable}, whose ISAC feasibility was validated in \cite{10438390}. The UEs are configured with heights $z_k=\SI{1.5}{m}$ and an antenna gain of \SI{3}{dBi} following \cite{9886383}. The RCS of the UAV is set to \SI{0.04}{m^2}, assuming a small drone as the sensing target \cite{patel2018review}. The noise power spectral density $\sigma_n^2 = \sigma_{n_k}^2$ is identical for all UEs, including noise figure of \SI{9}{dB} referring to \cite{andersson2017design}. The SNR threshold for strong received signal quality is set to \SI{20}{dB}, based on \cite{teltonikaSINR}. The CRB thresholds are selected to meet the high-accuracy demands of human living and working environments, as suggested in \cite{tan2021integrated}. The total power over the entire bandwidth $P_t$ is assigned to RIS-cascaded ISAC channels and BS2UAV sensing channel. Please notice that these parameters are generally chosen to represent worst-case conditions for RIS deployment, ensuring that the derived RIS size can support all possible conditions. The real-world environments is kept as the one in Fig.~\ref{RIS_ISAC_scenario}, yielding the area $\mathcal{R}_{1}^{\text{cov}}$ covers $A_1^{\text{cov}}=$\SI{4880}{m^2}, and the area $\mathcal{R}_{2}^{\text{cov}}$ covers $A_2^{\text{cov}}=$\SI{2347}{m^2}.

\begin{table}
\begin{center}
\caption{Simulation settings overview}
\label{sim_set}
\begin{tabular}{| c | c | c | c| }
\hline
Parameters & value & Parameters & value \\
\hline
$\boldsymbol{p}_b$ & $[45,8,30]$  & $PL_{\text{max}}$ &\SI{20}{dB}\\
\hline
$f_c$ & \SI{28}{GHz} & $\sigma_n^2$ & \SI{-165}{dBm/Hz} \\ 
\hline
$P_t$ & \SI{43}{dBm} & $B$ & \SI{1}{GHz}\\
\hline
$G_b$ & \SI{3}{dBi} & $N_c$ & 2560\\ 
\hline
$z_k$ & \SI{1.5}{m} & $M_{\text{OFDM}}$ & 2048\\ 
\hline 
$\alpha_u$ & \SI{0.04}{m^2} & $T_0$ & \SI{5.24}{ms}\\ 
\hline 
$M^{\text{ref}}$  & $20\times20$ & $d_{\min}$& \SI{0.3}{m}\\ 
\hline 
$z_u$ & \SI{50}{m} & $\gamma^{\text{tr}}$  &\SI{20}{dB}\\ 
\hline
$CRB_{d^{\text{tr}}}$& \SI{4e-4}{m^2} & $CRB_{v^{\text{tr}}}$ & \SI{1e-2}{(m/s)^2}\\ 
\hline
$N$& 2& $M_u$ & 25 \\ 
\hline
$A_1^{\text{cov}}$& \SI{4880}{m^2} & $A_2^{\text{cov}}$ & \SI{2347}{m^2}\\ 
\hline 
$K_1$& 54&  $K_2$ & 24\\ 
\hline 
$\eta_{\text{C}},\eta_{\text{S}}$& 0.3 &  $L$ & 2\\ 
\hline 
\end{tabular}
\end{center}
\end{table} 

\vspace{-0.5\baselineskip}
\subsection{RIS-ISAC Performance Evaluation} \label{ISAC_opt_results}
\begin{figure}[H]
\centering
\subfigure[]{\includegraphics[width=1.6in,trim=0mm 0mm 0mm 2mm,clip]{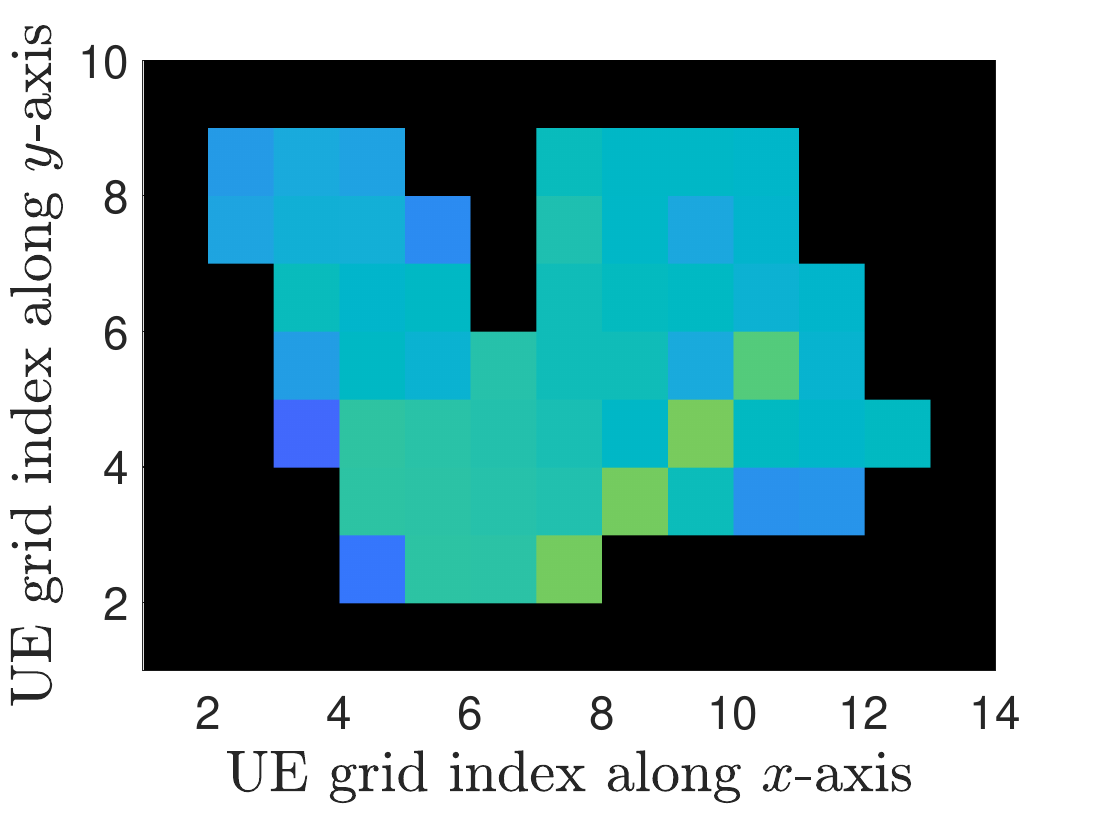}%
\label{only_bs_snr_high}
}
\hspace{-15pt}
\subfigure[]{\includegraphics[width=1.8in,trim=0mm 0mm 0mm 2mm,clip]{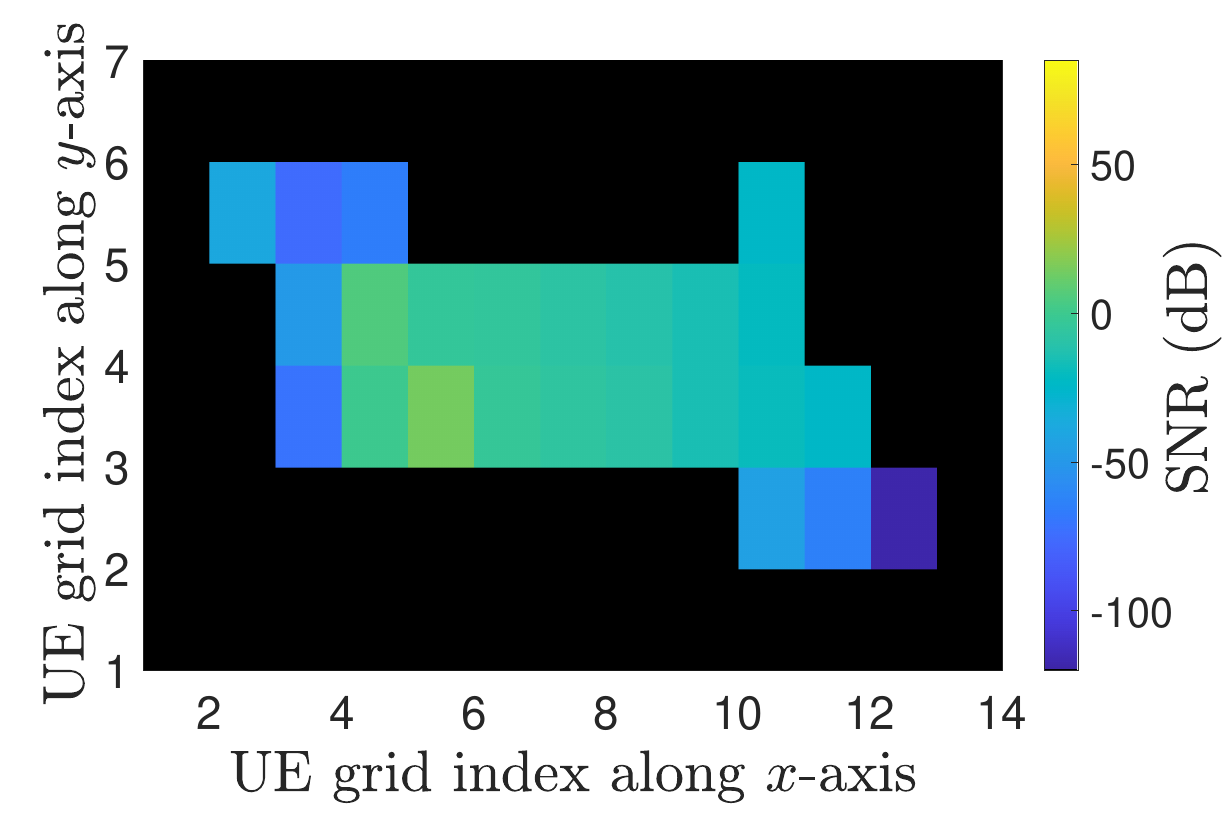}%
\label{only_bs_snr_low}
}
\caption{The SNR maps without RIS assistance for: (a) $\mathcal{R}_{1}^{\text{cov}}$, (b) $\mathcal{R}_{2}^{\text{cov}}$.}
\label{only bs UE SNR}
\end{figure}

Firstly, the communication signal coverage in SNR map format with $\gamma_{\text{C},b,k}$ from \eqref{Only BS SINR} at all the UE grids without RIS assistance is presented by Fig.~\ref{only_bs_snr_high} for $\mathcal{R}_{1}^{\text{cov}}$ and Fig.~\ref{only_bs_snr_low} for $\mathcal{R}_{2}^{\text{cov}}$. Using ARG method mentioned in Subsection \ref{RISposition}, both areas are divided into $10 \times$\SI{10}{m^2} grids, indexed as depicted in the figures. By evaluating the SNRs against the threshold $\gamma^{\text{tr}}$, the resulting coverage rates for $\mathcal{R}_{1}^{\text{cov}}$ and $\mathcal{R}_{2}^{\text{cov}}$ are both 0\%, with few grids above \SI{0}{dB} among the average value of \SI{-25}{dB}, and some grids even below \SI{-100}{dB}. This is primarily due to the lack of LoS access in the BS2UE links caused by building blockages. As a result, the signal propagation relies on reflection and diffraction paths, which suffer from severe losses at mmWave frequencies, further exacerbated by the wide bandwidth that increases noise power, ends up with low SNR, yielding the necessity of RIS assistance.

\begin{figure}[!t]
\centering
\includegraphics[width=2.5in,trim=50mm 55mm 0mm 0mm, clip]{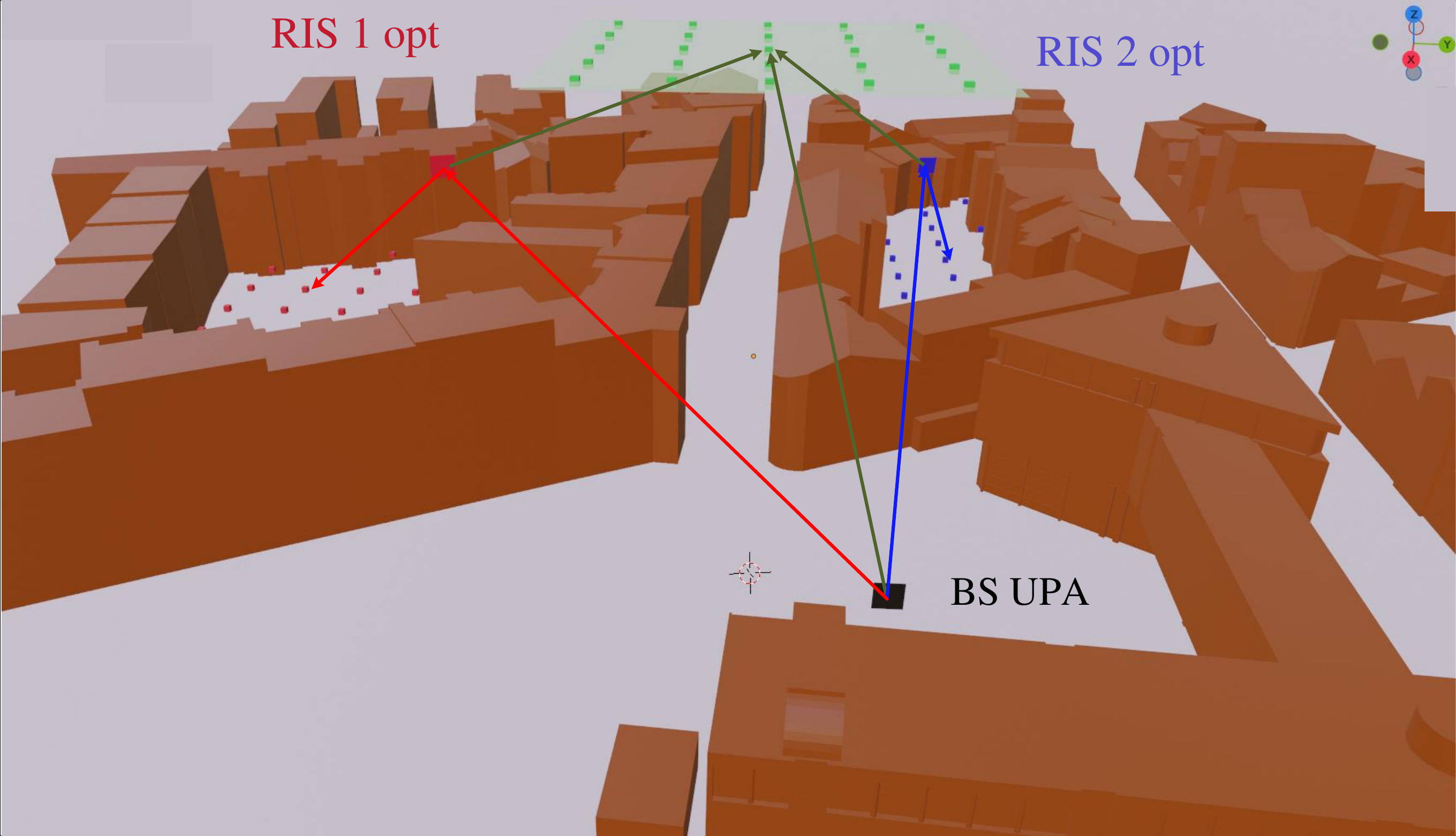}
\caption{The optimized RIS deployment for the $1$-st RIS (RIS 1 \textcolor{red}{\rule{6pt}{6pt}}) and the $2$-nd RIS (RIS2 \textcolor{blue}{\rule{6pt}{6pt}}). The cursor shows the coordinate origin.}
\label{ISAC_opt_pos}
\end{figure}

\begin{figure}[!t]
\centering
\subfigure[]{\includegraphics[width=1.7in,trim=0mm 0mm 0mm 0mm,clip]{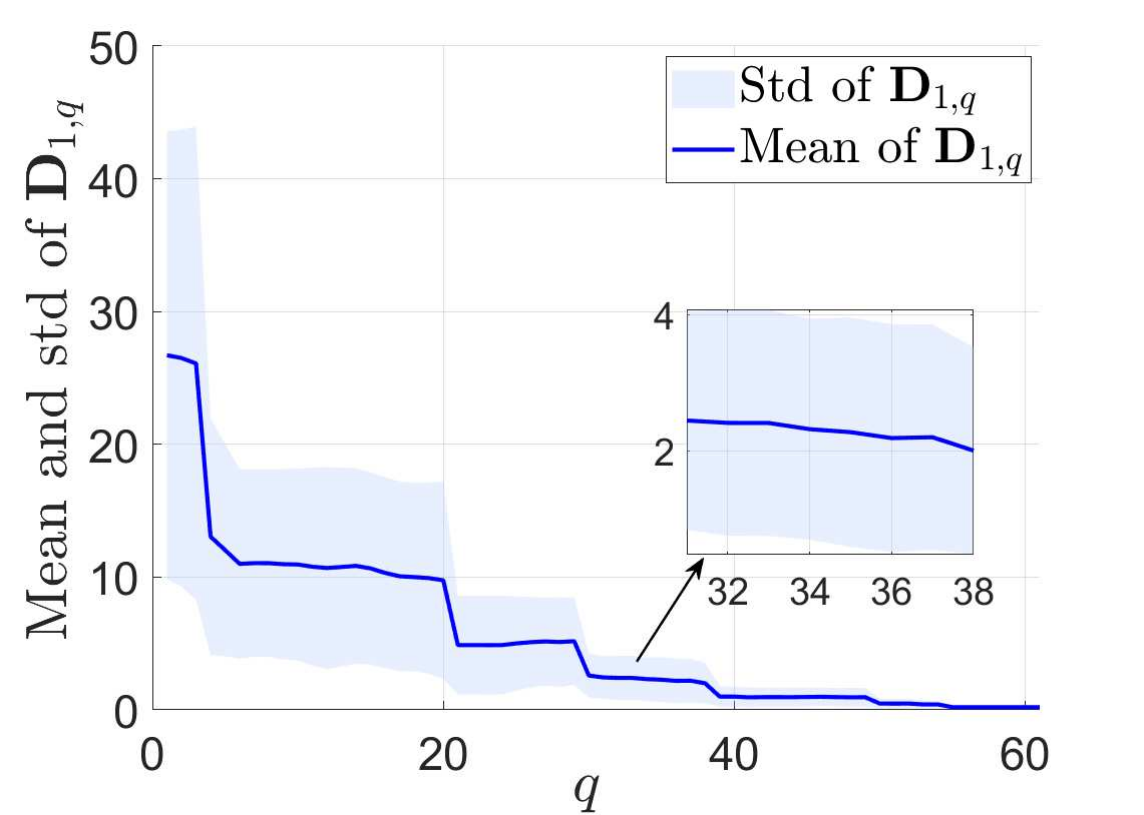}%
\label{dis_conv_ris_1}}
\hspace{-10pt}
\subfigure[]{\includegraphics[width=1.7in,trim=0mm 0mm 0mm 0mm,clip ]{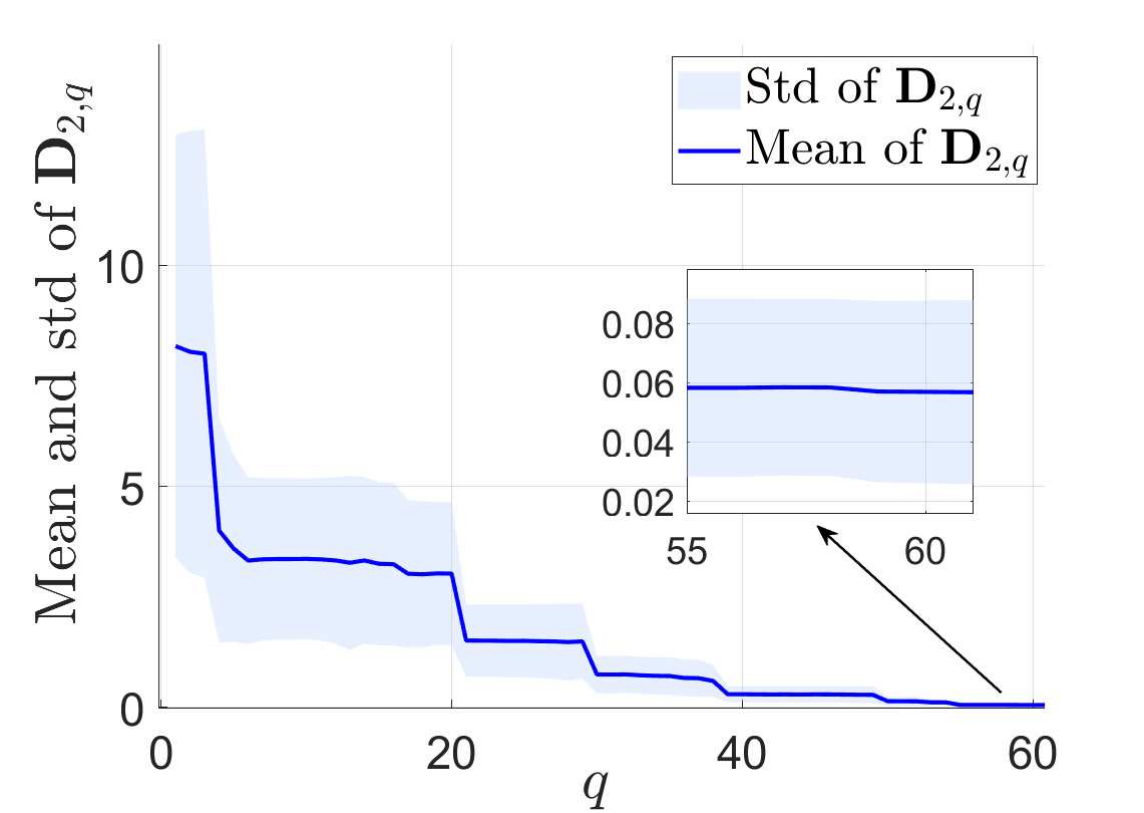}%
\label{dis_conv_ris_2}}
\caption{Convergence behavior of Alg.~\ref{algo_1} for: (a) RIS 1, (b) RIS 2.}
\label{n}
\end{figure}

\vspace{-0.2\baselineskip}
Considering the RIS deployment constraints in Subsection \ref{RISposition}, $N=2$ RISs are determined with each one serves its corresponding area $\mathcal{R}_{1}^{\text{cov}}$ and $\mathcal{R}_{2}^{\text{cov}}$. By solving \ref{prob:P1} with Alg.~\ref{algo_1}, the RISs are deployed as shown in Fig.~\ref{ISAC_opt_pos} with proportional appearance. The convergent behavior of Alg.~\ref{algo_1} is presented in Fig.~\ref{n}. Both the mean and standard deviation (std) of pairwise distance set $\textbf{D}_{n,q}$ in \eqref{conv_dist_set} decrease gradually as a general behavior as presented in the zoomed-in curve from Fig.~\ref{dis_conv_ris_1}. The step-wise behavior is due to the shrink operation of NM \cite{liu2024enhanced}. The std starts with large values to ensure the proper exploration of the entire RIS deployable regions as depicted in Fig.~\ref{RIS_ISAC_scenario}, which is also the reason why the values for RIS 1 are larger than RIS 2. It is visible in Fig.~\ref{dis_conv_ris_2} that over the last iterations, the mean value becomes stable, while $\max(\textbf{D}_{n,q})$ converges to $d_{\text{min}}$ after 61 iterations for both RISs. By then, the output of Alg.~\ref{algo_1} gives the RIS positions $\textbf{P}_{\text{opt}}= \{[-100.04,-58.50,18.66],[-142.19,36.55,10.06]\}$, orientations $\boldsymbol{\Theta}^{\text{r}}_{\text{opt}} = [12.94^{\circ}, 11.99^{\circ}]$ and $\boldsymbol{\Psi}^{\text{r}}_{\text{opt}}= [19.63^{\circ}, 6.33^{\circ}]$, sizes $\textbf{A}_{\text{opt}}= [4.55 \times 4.55,3.67\times 3.67]$ \SI{}{m^2} for the two RISs. 

\begin{figure}[!t]
\centering
\subfigure[]{\includegraphics[width=1.6in,trim=0mm 0mm 0mm 2mm,clip]{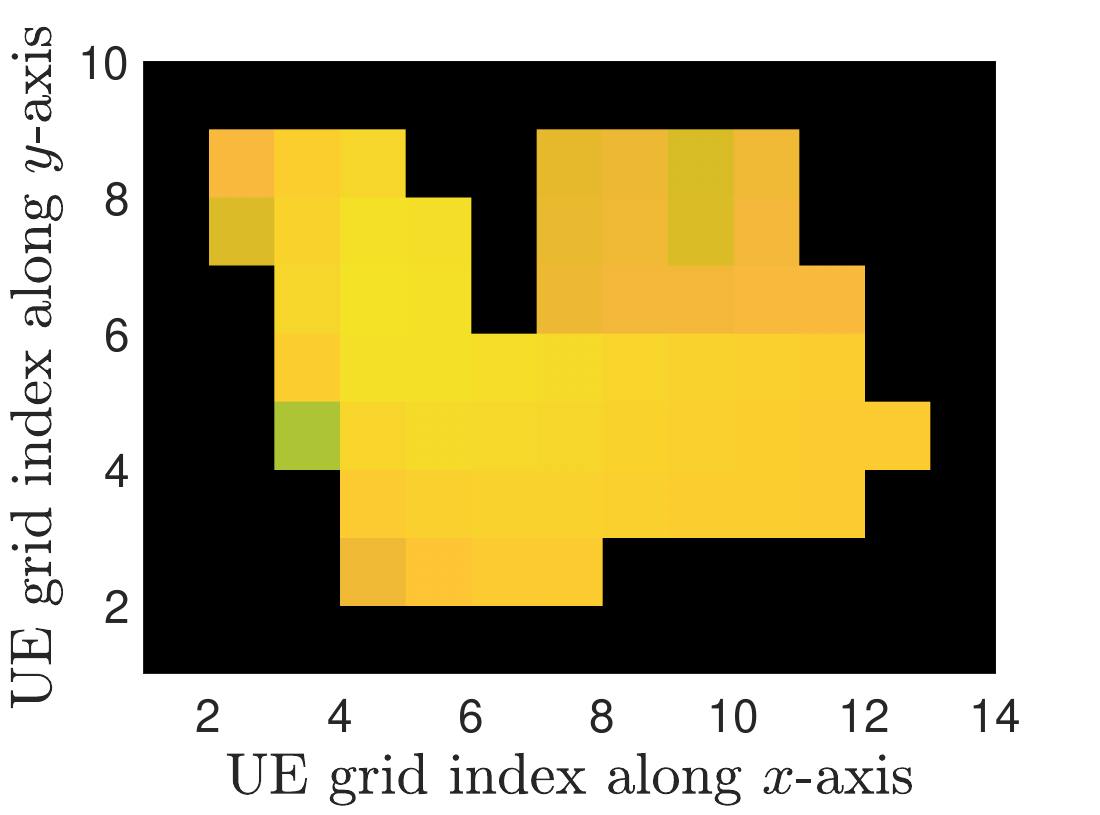}%
\label{ISAC_opt_area_1}}
\hspace{-15pt}
\subfigure[]{\includegraphics[width=1.8in,trim=0mm 0mm 0mm 2mm,clip]{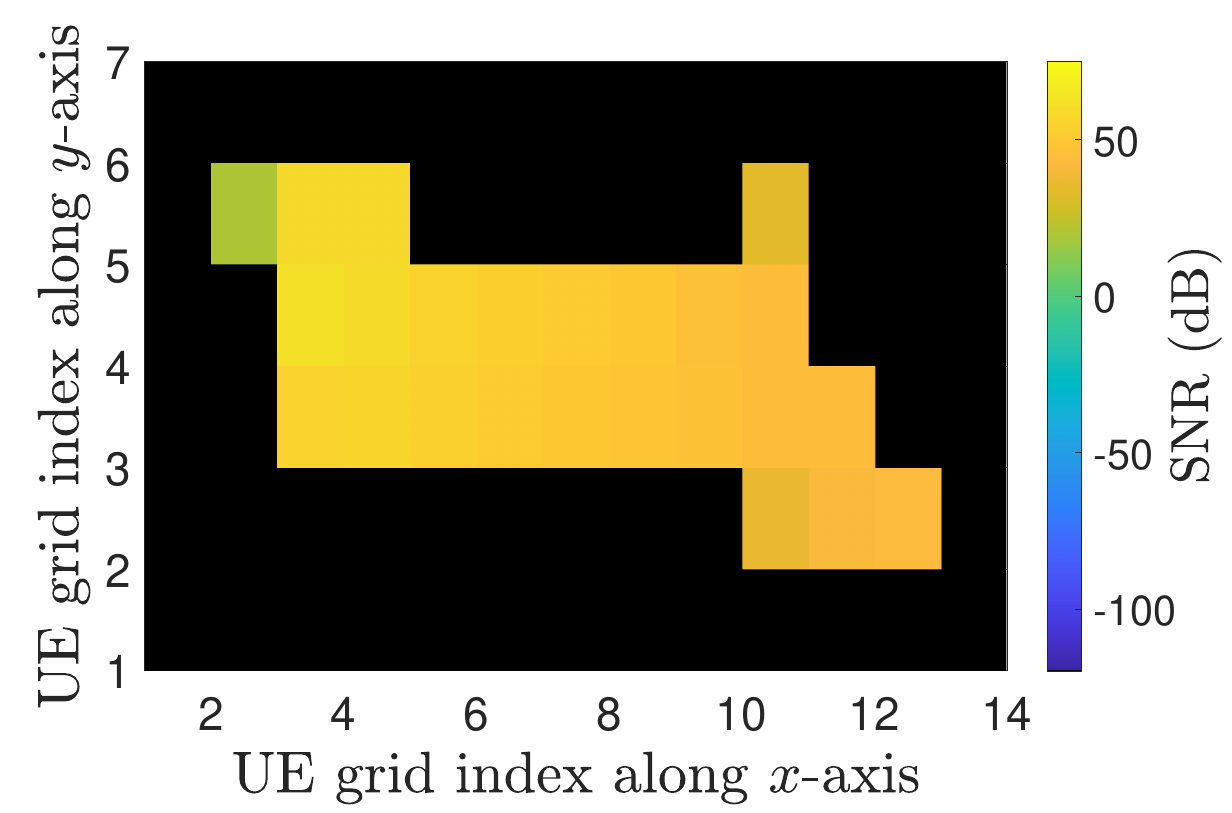}%
\label{ISAC_opt_area_2}}
\caption{The SNR maps with optimized RIS deployment for: (a) $\mathcal{R}_{1}^{\text{cov}}$, (b) $\mathcal{R}_{2}^{\text{cov}}$.}
\label{ISAC_opt_snr_min}
\end{figure}

\begin{figure}[!t]
\centering
\subfigure[]{\includegraphics[width=1.72in,trim=0mm 6mm 0mm 5mm,clip]{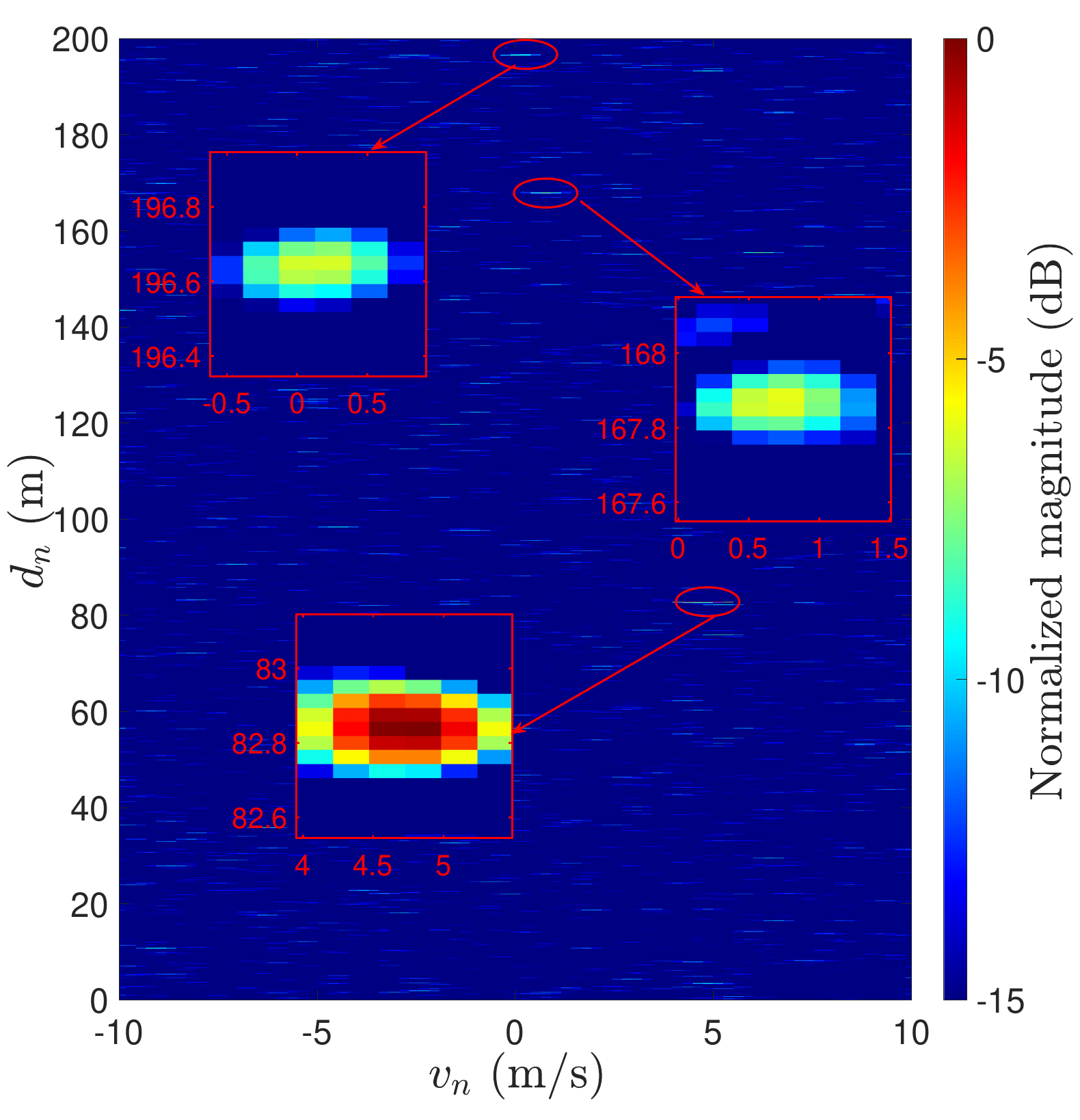}
\label{range_velocity_map}
}%
\hspace{-5pt}
\subfigure[]{\includegraphics[width=1.69in,trim=2mm 1mm 35mm 0mm,
    clip]{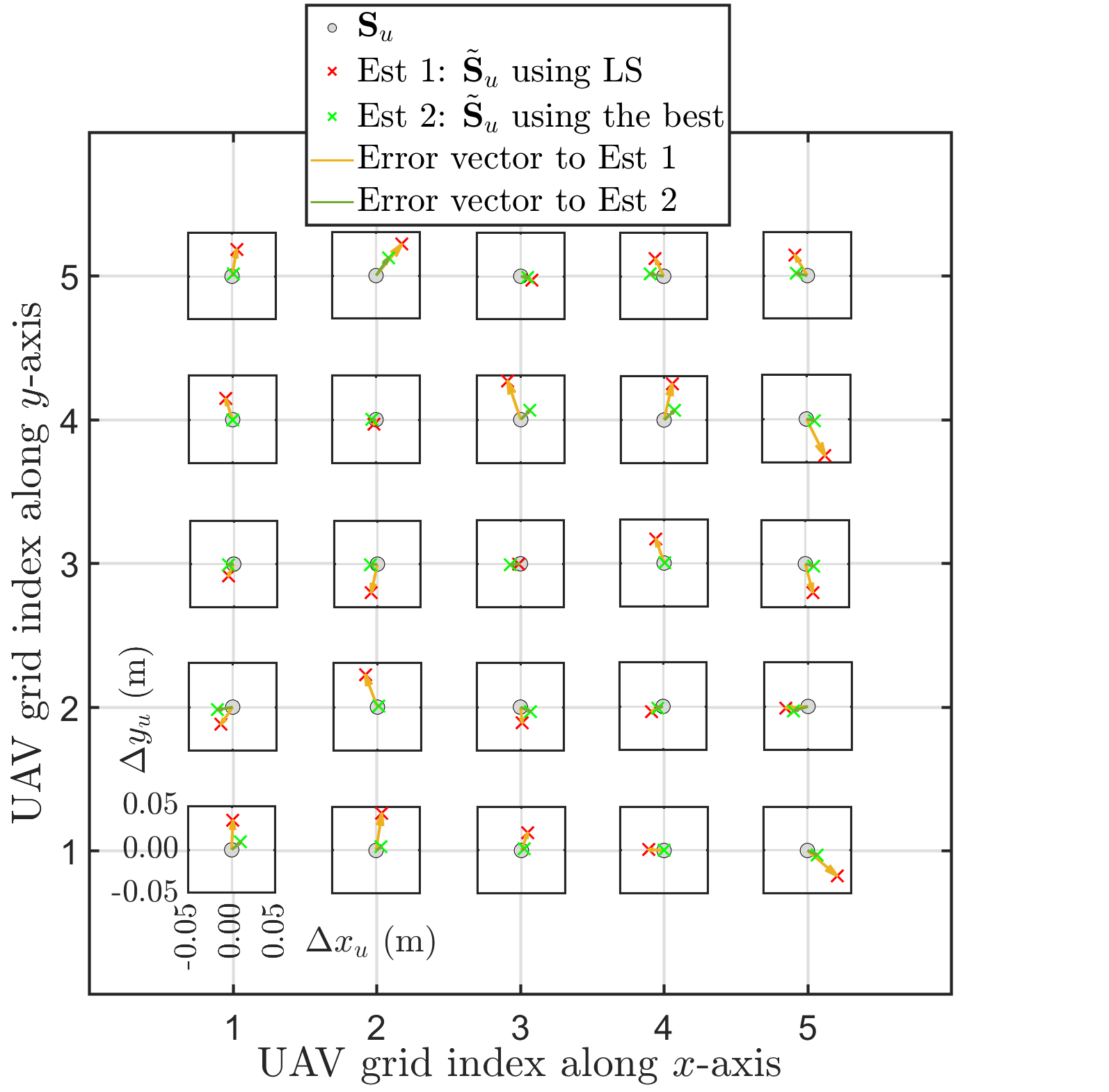}%
\label{sensing_position}
}
\caption{Sensing performance evaluation: (a) Range-velocity map. (b) Position estimation. }
\label{sensing results}
\end{figure}

As the UAV flies in the area $\mathcal{R}_{u}$, the RIS and BS beamforming weighting factors may change accordingly, to fulfill the ISAC QoS. To present the RIS-assisted areas are covered sufficient SNR mo matter where the UAV is, the SNR maps taking the minimum values among all the UAV positions are presented in Fig.~\ref{ISAC_opt_snr_min}. It could be obviously recognized from the colors that the communication coverage in the areas $\mathcal{R}_{1}^{\text{cov}}$ and $\mathcal{R}_{2}^{\text{cov}}$ are better than the case without RIS assistance in Fig.~\ref{only bs UE SNR}. In addition, the worse SNR is also larger than $\gamma^{\text{tr}}\!=\!20$ dB yielding a successful 100\% signal coverage. For sensing, the range-velocity map for an exemplary UAV at the middle of $\mathcal{R}_u$ flying along the $x$-axis with velocity \SI{5}{m/s} is presented. Although the map looks a bit noisy due to the CRB-orientated optimization, the three clusters can be still recognized using constant false alarm rate (CFAR) detector \cite{richards2010principles}, at $\tilde{\boldsymbol{\eta}}_0 = [$\SI{82.84}{m}, \SI{4.85}{m/s}$]^T, \tilde{\boldsymbol{\eta}}_1 = [$\SI{196.69}{m}, \SI{0.26}{m/s}$]^T,\tilde{\boldsymbol{\eta}}_2 = [$\SI{167.87}{m}, \SI{0.77}{m/s}$]^T$, indicating the BS2UAV, RIS 1 assisted, RIS 2 assisted sensing results for range and velocity, respectively. Utilizing these results to process multi-path sensing, the real UAV position set $\textbf{S}_u$ and the estimated set $\tilde{\textbf{S}}_u$ are presented in Fig.~\ref{sensing_position}, with $\Delta x_u$ and $\Delta y_u$ showing the position mismatch. For position estimation with least-square (LS) algorithm \cite{de2020range}, an average mismatch \SI{0.041}{m} is achieved. By leveraging the best range estimation, the position mismatch could be further reduced to \SI{0.01}{m}. Both results are under the effects of estimation resolution, but still could achieve the center-meter level accuracy mentioned in \cite{tan2021integrated}, proving the qualified sensing performance assisted by the RISs.

\begin{table*}
  \renewcommand{\arraystretch}{1.15}  
  \caption{Optimization results}
  \label{Compare}
  \centering
  \begin{tabular}{|c|c|c|c|c|c|c|}
    \hline
    \multirow{2}{*}{\textbf{ }}
      & \multicolumn{6}{c|}{\textbf{Properties Considered}} \\ \cline{2-7}
      & \textbf{Signal Model} & \textbf{Environment} 
      & \textbf{RIS Position} & \textbf{RIS Size}  & \textbf{RIS Orientation} & \textbf{Scalability}
      \\ \hline
   \cite{10172310} & Communication & Statistical & 1D position without constraint & \texttimes & \texttimes & Single RIS\\ \hline
   \cite{9687840} & Communication & Statistical  & 2D position without constraint & \texttimes & \texttimes & Single RIS\\ \hline
   \cite{yu2023active} & ISAC  & Statistical & 3D position withuot constraint & \texttimes & \texttimes & Single RIS\\ \hline
   \cite{liu2020ris} & Communication  & Real-world & 3D position practical constraint & \texttimes & \texttimes & Single RIS\\ \hline
   \cite{zeng2020reconfigurable} & Communication & Statistical  & 2D position without constraint & \texttimes & \checkmark & Single RIS\\ \hline
   \cite{ma2024multi} & Communication & Statistical  & 2D position without constraint & \texttimes & \texttimes & Multiple RIS\\ \hline
   \cite{fu2025multi} & Communication  & Real-world & 3D position without constraint & tile-wise size & \checkmark & Multiple RIS\\ \hline
    \cite{encinas2025isac} & ISAC  & Statistical & 3D position without constraint & tile-wise size & \texttimes & Single RIS\\ \hline
    This paper & ISAC  & Real-world & 3D position practical constraint & optimized size & \checkmark  & Multiple RIS\\ \hline    
  \end{tabular}
\end{table*}

\vspace{-0.5\baselineskip}
\subsection{Performance Comparisons}
A comparison between representative RIS deployment studies and the proposed solution in this paper is provided in Table~\ref{Compare}. For clarity, the position constraint refers to practical feasibility considerations in RIS deployment, such as building-attached deployment. The tile-wise size refers to increasing the RIS size with lower resolution, typically by adding or defining panels of fixed sizes. Overall, the proposed solution in this paper comprehensively incorporates environment,spatial, and signal processing features to multiple RISs in high-fidelity scenarios at mmWave frequencies, with the in-depth ISAC-driven modeling and analysis serving as a key innovation. The results of the analysis are summarized in Table~\ref{opt_results_tab}.

\subsubsection{Communication-Driven Case} 
\begin{figure}[!t]
\centering
\subfigure[]{\includegraphics[width=1.6in,trim=0mm 0mm 0mm 2mm,clip]{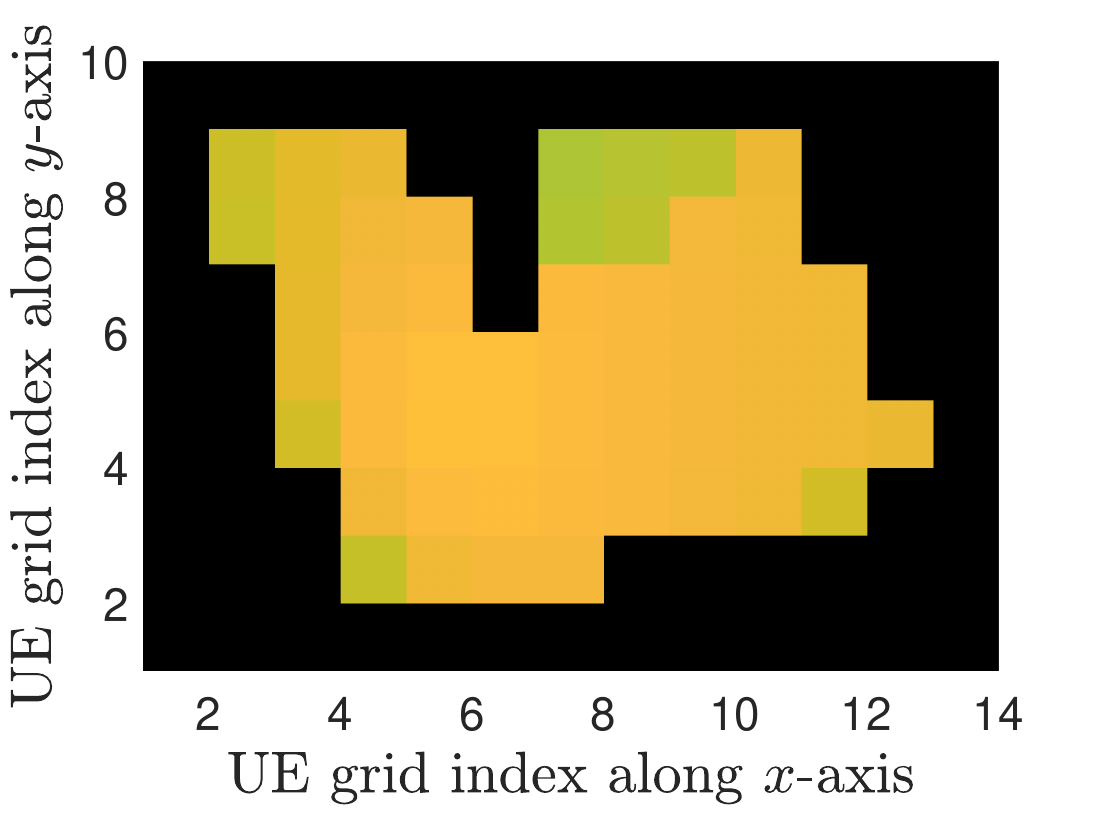}
}
\hspace{-15pt}
\subfigure[]{{\includegraphics[width=1.8in,trim=0mm 0mm 0mm 2mm,clip]{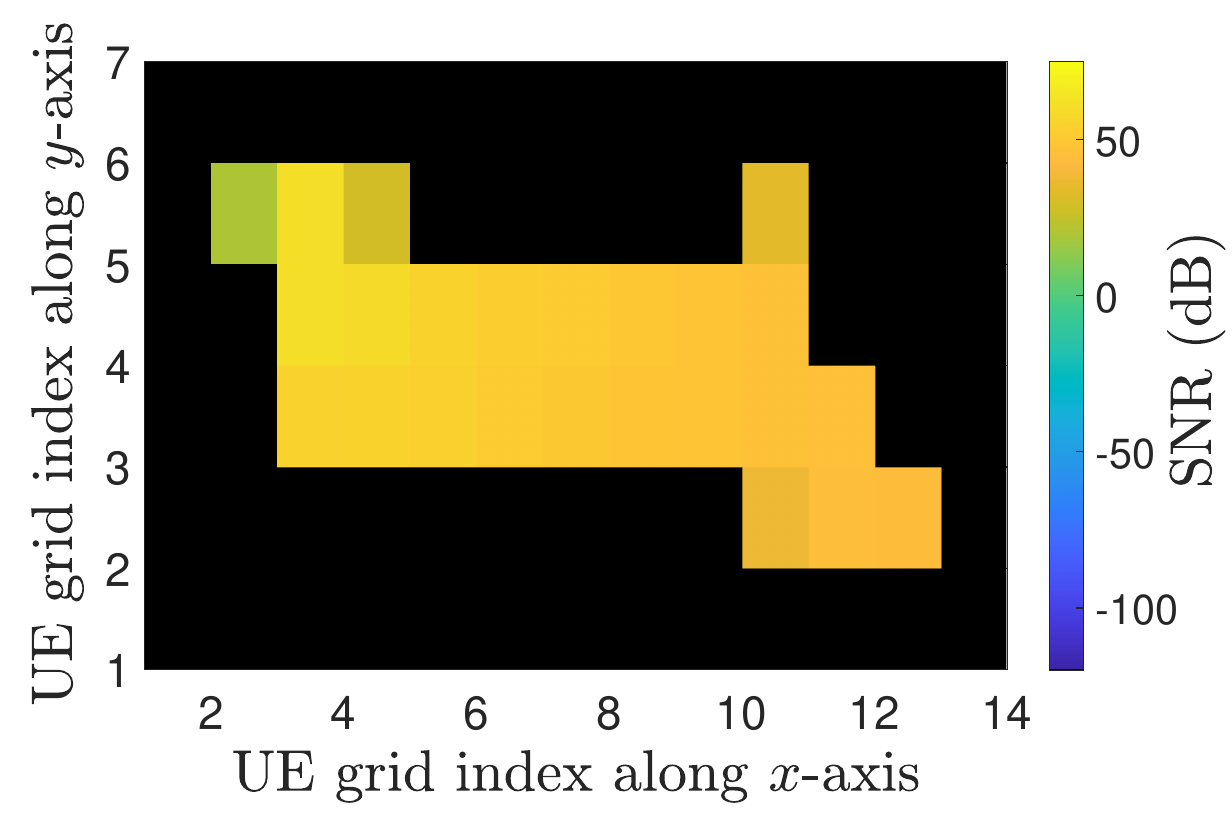}}%
}
\caption{The SNR map for communication-driven case: (a) $\mathcal{R}_{1}^{\text{cov}}$, (b) $\mathcal{R}_{2}^{\text{cov}}$.}
\label{Only_comm_results}
\end{figure}

The proposed algorithm also supports RIS deployment studies from Table~\ref{Compare}, where only communication QoS is considered, achieved by subtracting the sensing contribution in \ref{prob:P0} and Alg.~\ref{algo_1}. The output gives the positions of RISs $\textbf{P}_{\text{opt}}= \{[-91.85,-76.06,22.58],[-143.11,41.57,9.52]\}$, orientations $\boldsymbol{\Theta}^{\text{r}}_{\text{opt}} = [16.88^{\circ}, 18.35^{\circ}]$ and $\boldsymbol{\Psi}^{\text{r}}_{\text{opt}}= [353.88^{\circ}, 27.56^{\circ}]$, sizes $\textbf{A}_{\text{opt}}= [1.98 \times 1.98, 2.51\times 2.51]$ \SI{}{m^2}, realizing the UE SNR map as depicted in Fig.~\ref{Only_comm_results}. However, although the communication coverage is 100\% served, the sensing $CRB_{d^{\text{tr}}}$ and $CRB_{v^{\text{tr}}}$ can not be satisfied, even if RIS reflects all the power to UAV yielding $\beta_{n,u}=1$. This highlights the necessity of considering ISAC properties when deploying RIS to ensure the effectiveness of future dual-functional systems.
\subsubsection{Path-Loss-Driven Method} 
RIS deployment could be also realized by looking for the position with the minimum FSPL as \cite{10172310}, \cite{9687840} and \cite{yu2023active}. Following the same procedure, the RIS positions are determined at $\textbf{P}_{\text{opt}}= \{[-90.60,-66.60,18.20],[-133.80,27.41,11.47]\}$. However, even with optimum orientations $\boldsymbol{\Theta}^{\text{r}}_{\text{opt}} = [15.12^{\circ}, 18.64^{\circ}]$ and $\boldsymbol{\Psi}^{\text{r}}_{\text{opt}}= [23.80^{\circ}, 26.82^{\circ}]$ derived from Subsection \ref{fixed_pos_algo1} applied, the RIS sizes yields $\textbf{A}_{\text{opt}}= [10.50 \times 10.50, 7.19\times 7.19]$ \SI{}{m^2} to satisfy the required ISAC QoS, which is outperformed by the proposed solution in this paper. This is caused by the reason that the FSPL computation can not represent all the physical features of the environments, yielding sub-optimum solutions.
\subsubsection{Deployment with Passive Orientation} 
\begin{figure}[!t]
\centering
\subfigure[]{\includegraphics[width=1.6in, trim = 0mm 0mm 0mm 2mm,clip]{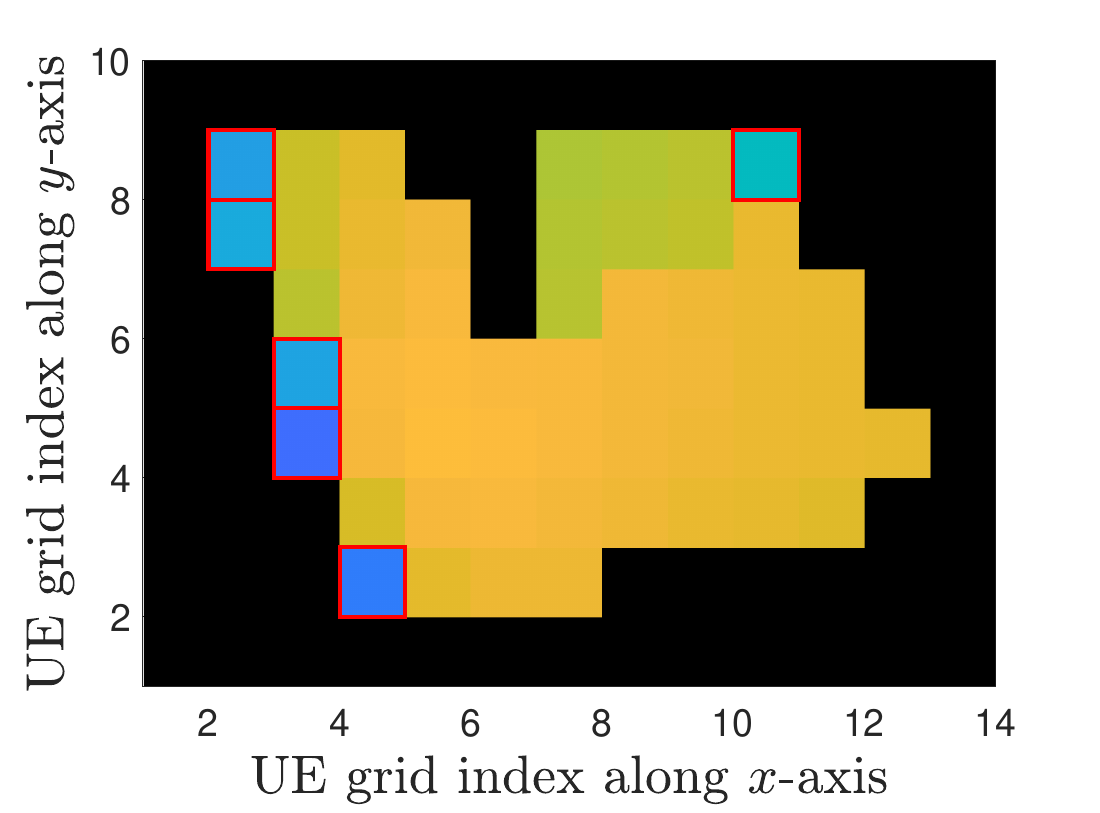}
\label{ori_attach_to_building_high_f}}
\hspace{-15pt}
\subfigure[]{\includegraphics[width=1.8in, trim = 0mm 0mm 0mm 2mm,clip]{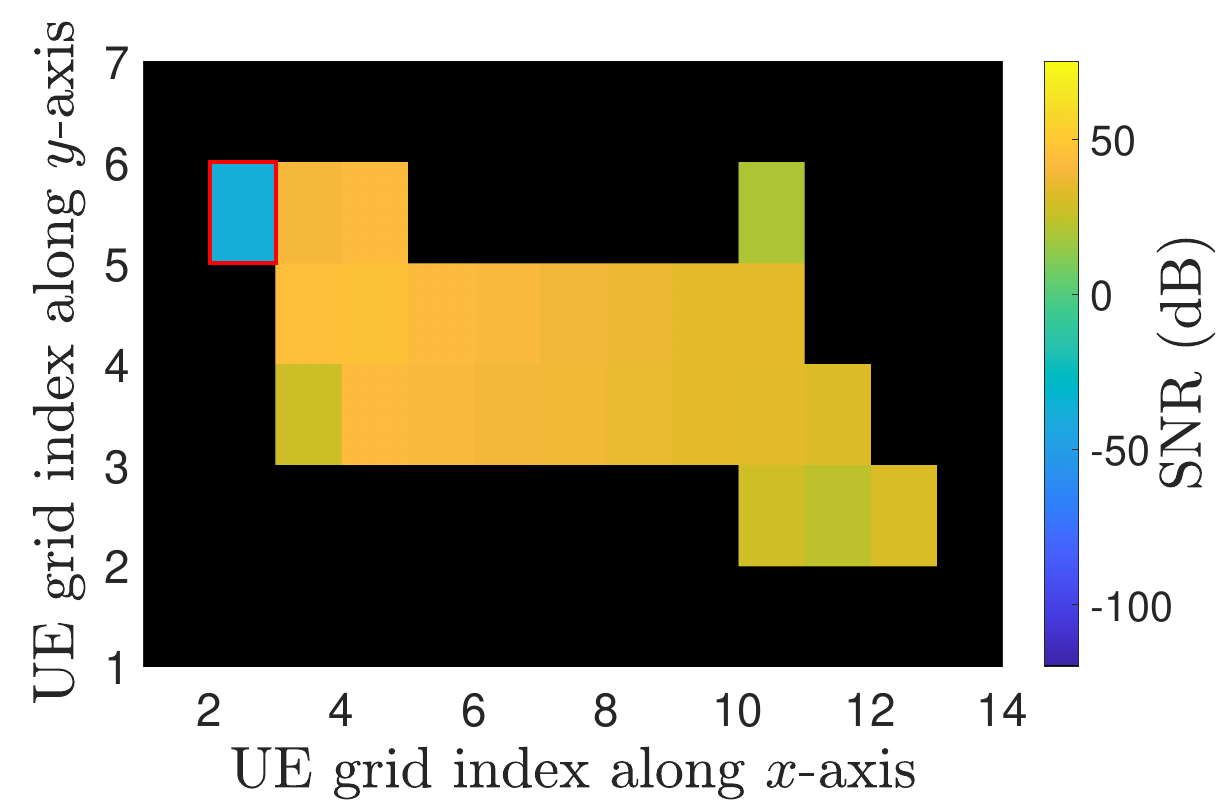}%
\label{ori_attach_to_building_low_f}}
\caption{The SNR map with passive orientation for: (a)$\mathcal{R}_{1}^{\text{cov}}$, (b) $\mathcal{R}_{2}^{\text{cov}}$. The uncovered grids are filled with BS2UE SNR in Fig.~\ref{only bs UE SNR}.}
\label{ori_attach_to_building}
\end{figure}

Following the approach of \cite{liu2020ris} and \cite{ma2024multi}, the RIS deployment with a fixed orientation is analyzed, referred to as passive orientation, without considering the movable antenna concept \cite{zhu2023modeling}. In the proposed environment, this results in $\boldsymbol{\Theta}^{\text{r}}_{\text{opt}} = [0^{\circ}, 0^{\circ}]$ and $\boldsymbol{\Psi}^{\text{r}}_{\text{opt}} = [18^{\circ}, 22^{\circ}]$, aligned with the building sidewall. A key drawback of this method is that the reachability of the RIS-assisted area is not fully achieved, leading to uncovered areas, as indicated by the red-edged grids in Fig.~\ref{ori_attach_to_building}. Nevertheless, \ref{prob:P1} could be still solved by Alg. \ref{algo_1} with incomplete $\mathcal{R}_{1}^{\text{cov}}$ and $\mathcal{R}_{2}^{\text{cov}}$, resulting in RIS positions $\textbf{P}_{\text{opt}} = \{[-137.50,31.00,9.50],[-95.50,-70.85,17.25]\}$ and sizes $\textbf{A}_{\text{opt}} = [2.93\times 2.93, 3.22 \times 3.22]$ \SI{}{m^2}, as shown in the SNR map in Fig.~\ref{ori_attach_to_building}, with coverage ratios of $88.89\%$ and $95.83\%$, respectively. The grids that are not covered by the RIS have to rely on the BE2UE link, leading to severe quality degradation, thus undermining the purpose of RIS deployment. While additional RIS deployment elsewhere could address this issue, the increased hardware cost would conflict with the energy efficiency goal of minimizing the multi-RIS size-to-coverage sum ratio.

\subsubsection{RIS with 1-bit Phase Resolution} 
Since RIS with $L=1$ bit phase resolution has the lowest-cost and easiest-to-fabricate advantages, the deployment of such design is also investigated. In this case, Alg.~\ref{algo_1} outputs the positions $\textbf{P}_{\text{opt}}= \{[-100.04,-58.50,18.66],[-142.19,36.55,10.06]\}$, orientations $\boldsymbol{\Theta}^{\text{r}}_{\text{opt}} = [12.94^{\circ}, 11.99^{\circ}]$ and $\boldsymbol{\Psi}^{\text{r}}_{\text{opt}}= [19.63^{\circ}, 6.33^{\circ}]$, and sizes $\textbf{A}_{\text{opt}}= [5.41 \times 5.41,4.12\times 4.12]$ \SI{}{m^2}. The major change is the RIS size, since the reduced phase resolution from 2-bit to 1-bit mainly results in the effective antenna gain reduction as proved in \cite{li2024intelligent}.

In summary, the proposed algorithm in this paper successfully satisfies the ISAC QoS with optimized RIS deployment, whereas previous algorithms suffer from issues such as larger RIS sizes, insufficient communication coverage, unsupported sensing capabilities, or mismatching real-world environments, which clarifies the necessity and impact of RIS-ISAC deployment studies in
real-world environments.

\begin{table*}[t!]
  \renewcommand{\arraystretch}{1.15}  
  \caption{Optimization results}
  \label{opt_results_tab}
  \centering
  \begin{tabular}{|c|c|c|c|c|c|c|}
    \hline
    \multirow{2}{*}{\textbf{ }}
      & \multicolumn{3}{c|}{\textbf{RIS 1}} 
      & \multicolumn{3}{c|}{\textbf{RIS 2}} \\ \cline{2-7}
      & \textbf{Size} & \textbf{UE Coverage}  & \textbf{Sensing Ability}
      & \textbf{Size} & \textbf{UE Coverage}  & \textbf{Sensing Ability} \\ \hline
   Proposed algorithm &$4.55 \times 4.55$\SI{}{m^2}& $100\%$ & satisfied & $3.67 \times 3.67$\SI{}{m^2} & $100\%$ & satisfied\\ \hline
   Only Communication &$1.98 \times 1.98$\SI{}{m^2}& $100\%$ & \textcolor{red}{Not available} & $2.51 \times 2.51$\SI{}{m^2} & $100\%$ & \textcolor{red}{Not available} \\ \hline
   Path-loss based &\textcolor{red}{$10.50 \times10.50$\SI{}{m^2}}& $100\%$ & satisfied & \textcolor{red}{$7.19 \times 7.19$\SI{}{m^2}} & $100\%$ & satisfied\\ \hline
   Passive Orientation &$2.93 \times 2.93$\SI{}{m^2}& \textcolor{red}{ $88.89\%$} & satisfied & $3.22 \times 3.22$\SI{}{m^2} & \textcolor{red}{$95.83\%$} & satisfied\\ \hline
   Proposed algorithm ($L=1$) &$5.41 \times 5.41$\SI{}{m^2}& $100\%$ & satisfied & $4.12 \times 4.12$\SI{}{m^2} & $100\%$ & satisfied\\ \hline
  \end{tabular}
\end{table*}

\vspace{-0.5\baselineskip}
\section{Conclusion and Outlook}
\label{sec:conc}
This paper proposes an optimization framework for multi-RIS deployment of mmWave ISAC systems in real-world environments. Based on ISAC signal modeling using IHE-RL, an energy-efficiency-driven problem is formulated to minimize the multi-RIS size-to-coverage sum ratio, considering practical deployment constraints, positions, orientations, and beamforming strategies at both the BS and the RISs. The problem is simplified via an equivalent gain scaling method, then solved using a two-step iterative approach. Simulation results with realistic parameters demonstrate that the proposed strategy effectively achieves full dual-functionality with reduced RIS size by incorporating a more comprehensive and practical deployment design. Future work includes building a dataset using the proposed method to support AI-driven algorithms, and applying more standardized configurations once 6G specifications are determined for ISAC, RIS, and beamforming strategies.

\vspace{-0.5\baselineskip}
\appendix[Calculation of CRB] \label{Appendix_part}
Based on the signal model in \eqref{SensingsignalwithRIS2}, the FIM for estimating $\boldsymbol{\eta}_n$ is given by \cite{li2008mimo} 
\begin{equation}\label{FIM_time}
    \textbf{J}_{\boldsymbol{\eta}_n}=\frac{2}{\sigma_{n_k}^{2}} \Re\left\{\int_{T_{0}} \frac{\partial(\boldsymbol{h}_{e}^{\text{tot}}(t, \boldsymbol{\eta}_n) \alpha_n)^{H}}{\partial \boldsymbol{\eta}_n} \frac{\partial(\boldsymbol{h}_{e}^{\text{tot}}(t,\boldsymbol{\eta}_n) \alpha_n)}{\partial \boldsymbol{\eta}_n} d t\right\}.
\end{equation}
Using the definition of $\boldsymbol{h}_{e}^{\text{tot}}(t,\boldsymbol{\eta}_n)$ and considering $d_n = c_0\tau_n/2$ and $v_n = \lambda_c f_{D_n}/2$, the derivatives required for calculation of \eqref{FIM_time} are given by 
\begin{align}
\frac{\partial \boldsymbol{h}_{e}^{\text{tot}}(t, \boldsymbol{\eta}_n) \alpha_n}{\partial d_n} 
&= -\frac{2 \alpha_n}{c_0}  \dot{s}_{k}(t - \tau_n)  e^{j 2\pi f_{D_n} t}, \\
\frac{\partial \boldsymbol{h}_{e}^{\text{tot}}(t, \boldsymbol{\eta}_n)  \alpha_n}{\partial v_n} 
&= \frac{j4\pi t \alpha_n}{\lambda_c}  s_{k}(t - \tau_n) e^{j 2\pi f_{D_n} t},
\end{align}
where $\dot{s}_{k}(t)= d s_{k}(t)/d t$.
Therefore, the elements of the FIM matrix can be calculated based on \eqref{FIM_time}
\begin{equation} \label{FIM_1}
J_{d_n,d_n} = \frac{8|\alpha_n|^2}{\sigma_{n_k}^2 c_0^2} \Re\left\{ \int_0^{T_0} \dot{s_{k}}(t - \tau_n)  \dot{s_{k}}^H(t - \tau_n)  \, dt\right\},
\end{equation}
\begin{equation}
J_{v_n,d_n} =\frac{16\pi|\alpha_n|^2}{\sigma_{n_k}^2\lambda_c c_0} \, \Re \left\{ j
\int_0^{T_0}  t  \dot{s}_{k}(t - \tau_n) \, s_{k}^H(t - \tau_n) \, \mathrm{d}t \right\},
\end{equation}
\begin{equation} \label{FIM_3}
J_{v_n,v_n} = \frac{32\pi^2|\alpha_n|^2}{\sigma_{n_k}^2\lambda_c^2} \Re \left\{
\int_0^{T_0} t^2 s_{k}(t - \tau_n) \, s_{k}^H(t - \tau_n) \, \mathrm{d}t \right\}.
\end{equation}
Finally, all those terms can be put into \eqref{FIM_total} and CRBs can be calculated using \eqref{SensingRISCRB_d} and \eqref{SensingRISCRB_v}. All the FIM elements from \eqref{FIM_1} to \eqref{FIM_3} include the term $\frac{|\alpha_n|^2}{\sigma_{n_k}^2}$, which can be regarded as the SNR of the received sensing signal for the $n$-th path, resulting in the calculated CRBs inversely proportional to it.

\bibliography{mybibliography.bib} 
\bibliographystyle{ieeetr} 

\end{document}